\documentclass[manuscript, screen]{acmart}

\AtBeginDocument{%
  }

\setcopyright{acmlicensed}
\copyrightyear{2025}
\acmYear{2025}
\acmDOI{XXXXXXX.XXXXXXX}

\acmISBN{978-1-4503-XXXX-X/18/06}

\usepackage{algorithmic}
\usepackage{graphicx}
\usepackage{textcomp}
\usepackage{xcolor}
\usepackage{xspace}
\usepackage{booktabs}
\usepackage{subcaption}
\usepackage{graphicx}
\usepackage{multirow} 
\usepackage{url}
\usepackage{balance}
\usepackage{tabularx}
\usepackage{enumerate}
\usepackage[shortlabels]{enumitem}
\usepackage[T1]{fontenc}
\usepackage{comment}
\usepackage{amsmath}
\usepackage{mathtools}
\usepackage{siunitx}
\usepackage{epstopdf}
\usepackage{psfrag}
\usepackage{booktabs} 
\usepackage{threeparttable} 

\newcommand{\fakepar}[1]{\vspace{1.00mm}\noindent{#1}}
\newcommand{\boldpar}[1]{\vspace{1.50mm}\noindent\textbf{#1}}
\newcommand{\italicpar}[1]{\vspace{1.00mm}\noindent\textit{#1}}

\newcommand{\highlight}[1]{\color{black}#1\color{black}\xspace}

\newcommand{\NAME}{\mbox{\textsc{APEX}}\xspace}

\begin{document}

\title{\NAME: Automated Parameter Exploration for Low-Power Wireless Protocols}

\author{Mohamed Hassaan M. Hydher}
\email{m.h.mohamedhyder@tugraz.at}
\orcid{0000-0002-1685-9122}
\affiliation{%
  \institution{Graz University of Technology}
  \city{Graz}
  \country{Austria}
}

\author{Markus Schu{\ss}}
\email{markus.schuss@tugraz.at}
\orcid{0000-0002-8651-3725}
\affiliation{%
  \institution{Graz University of Technology}
  \city{Graz}
  \country{Austria}}

\author{Olga Saukh}
\email{saukh@tugraz.at}
\orcid{0000-0001-7849-3368}
\affiliation{%
  \institution{Graz University of Technology \& Complexity Science Hub}
  \city{Graz/Vienna}
  \country{Austria}
  }

\author{Kay R\"{o}mer}
\email{roemer@tugraz.at}
\orcid{0000-0002-4248-4424}
\affiliation{%
  \institution{Graz University of Technology}
  \city{Graz}
  \country{Austria}}

\author{Carlo Alberto Boano}
\email{cboano@tugraz.at}
\orcid{0000-0001-7647-3734}
\affiliation{%
  \institution{Graz University of Technology}
  \city{Graz}
  \country{Austria}\vspace{+5.00mm}}

\renewcommand{\shortauthors}{H. Hydher et al.}


\begin{abstract} 
Careful parametrization of networking protocols is crucial to maximize the performance of low-power wireless systems and ensure that stringent application requirements can be met. This is a non-trivial task involving thorough characterization on testbeds and requiring expert knowledge. Unfortunately, the community still lacks a tool to facilitate parameter exploration while minimizing the necessary experimentation time on testbeds. Such a tool would be invaluable, as exhaustive parameter searches can be time-prohibitive or unfeasible given the limited availability of testbeds, whereas non-exhaustive unguided searches rarely deliver satisfactory results. 
In this paper, we present APEX, a framework enabling an automated and informed parameter exploration for low-power wireless protocols and allowing to converge to an optimal parameter set within a limited number of testbed trials. We design APEX using Gaussian processes to effectively handle noisy experimental data and estimate the optimality of a certain parameter combination. After developing a prototype of APEX, we demonstrate its effectiveness by parametrizing two IEEE 802.15.4 protocols for a wide range of application requirements. Our results show that APEX can return the best parameter set with up to 10.6x, 4.5x and 3.25x less testbed trials than traditional solutions based on exhaustive search, greedy approaches, and reinforcement learning, respectively.

\end{abstract}




\maketitle


\section{Introduction} 
\label{sec:introduction} 

\noindent
Low-power wireless (LPW) systems have become an integral part of the Internet of Things (IoT) and lay the foundations for a wide spectrum of applications that are of utmost importance for our society. 
Such applications include smart and efficient buildings, precision agriculture, smart health, asset tracking, and smart manufacturing -- all very different in their nature and performance requirements. 
For~example, in smart farming systems, it is often important to maximize energy efficiency, so to minimize the need to replace batteries and increase user acceptance. 
On~the contrary, in industrial IoT systems, high reliability and short delays are pivotal to ensure correct operation and quickly detect anomalies. 
As~there is no ``one-size-fits-all'' solution in the IoT realm, a large number of LPW communication technologies and protocols have been proposed to satisfy largely different application requirements. 
These LPW solutions provide support for highly-diverse network topologies 
and medium access control (MAC) strategies -- thereby giving developers the chance to customize to the fullest extent their system to the application at hand in order to maximize performance~\cite{foukalas19dependability}.

\boldpar{Challenges.} Customizing a system for a specific application involves picking the right networking stack and configuring its protocols to maximize performance: unfortunately, this is not a trivial task, as several challenges need to be tackled. 

\italicpar{Expert knowledge required.} 
LPW protocols often rely on a set of parameters that must be chosen meticulously, \highlight{and their choice is not straightforward. For instance, consider the radio's transmission power parameter ($tx\_power$) in a LPW protocol. This parameter directly influences both energy consumption and packet reception ratio (PRR). The relationship between $tx\_power$ and PRR can be complex. While increasing $tx\_power$ may initially improve PRR, it can also lead to excessive interference in dense networks, ultimately reducing PRR in certain conditions. Similarly, one might intuitively expect the overall energy consumption to increase with higher $tx\_power$; however, this may not necessarily be the case 
(as we will illustrate in Fig.~\ref{Baloo_scatter}). 
This complexity makes it hard for users to identify the optimal 
parameters for a given application without a comprehensive understanding of these intricate dynamics. Moreover, the performance of MAC protocols can vary drastically as a function of the employed parameters, as it was shown in the context of several LPW technologies~\cite{spoerk17bleach,salomon24adaptive,Schub2018,grosswindhager18uwbparameters,schuh24complex,bor17lora,cattani17jsan}.} 
While default settings for each parameter exist, they are commonly meant for the general case, and are often sub-optimal for a given application~\cite{schuss17competition}. A careful configuration and customization of LPW protocols to the application at hand is hence a must, but it represents a complex and tedious task that requires specialized expertise and that strongly depends on aspects such as the employed platform, network topology and stack, as well as the type and amount of transmitted data. Therefore, in order to properly understand the role of each parameter, to quantitatively assess whether certain application requirements can be met, and to compare the performance of different configurations, one commonly needs to also gather experimental data. However, this is also not trivial, as discussed next. 

\italicpar{Experimentation is expensive.}
\highlight{
Parameter optimization ideally takes place directly in the actual deployment scenario. 
However, testing and optimizing a solution at the deployment site is typically not viable, due to the high costs and labor involved, as well as due to the inability to perform repeatable and controlled experiments over a prolonged time. As a result, simulation platforms and testbed facilities are often used to test and configure the developed solutions in more convenient settings and in environments that mimic the conditions of real-world deployments.} Simulation platforms for LPW systems such as COOJA~\cite{COOJA}, OMNeT++~\cite{OMNeT}, and \textit{ns-3}~\cite{NS_3_survey} allow fully-controllable and repeatable experiments, thus enabling parameter exploration without the need for expensive field trials. \highlight{However, these simulators often fail to capture key environmental factors, such as temperature fluctuations, RF interference, and hardware variability, which can significantly impact real-world performance\footnote{Notably, simulation results can deviate by up to 50\% compared to real-world tests performed on testbeds~\cite{Ugo_2007}.}. 
In contrast, LPW testbeds, such as FIT IoT-LAB~\cite{adjih15iotlab}, Indriya~\cite{Appavoo2019}, and D-Cube~\cite{Schub2018}, enable the evaluation of protocol performance on actual hardware (HW) in real-world settings, and may even allow to control environmental conditions such as RF interference~\cite{schuss19jamlabng} and temperature variations~\cite{boano14templab}. 
For this reason, experimentation on LPW testbeds is typically preferred~\cite{boano_18}.}
However, testbeds \highlight{--~especially public facilities~--} are often shared by several users, and their availability (i.e., the usable experimentation time) may hence be limited, \highlight{e.g., one might only get one hour of experimentation time per day}. 
This makes an exhaustive exploration of all possible parameters -- or an exploration giving a sufficient statistical significance -- impractical; especially considering that testbed runs may need to have a minimum duration to allow the network to setup and stabilize~\cite{banaszek24rpl}, \highlight{or that certain metrics (such as the PRR) may require the exchange of a large number of packets to be computed with sufficient granularity. 
For instance, when a node transmits only 20 packets within a testbed run, a receiver can compute the corresponding PRR only with a granularity of 5\%.} 

\italicpar{Lack of adequate tools.}
Unfortunately, to date, the community still lacks a simple tool 
to automatically parametrize LPW protocols based on the requirements of a given application. 
Existing work has proposed frameworks requiring extensive testbed usage (e.g., by exhaustively testing all parameter combinations~\cite{Fu2015}), demanding a deep understanding of the protocol internals~\cite{Zlmmerling2014}, or requiring the user to specify complex models of the environment and the employed hardware, which can be daunting and error-prone for non-experts~\cite{Oppermann2015}. 

\boldpar{Our contributions.} 
In this work, we introduce \NAME, a modular framework enabling an automated and efficient 
parameter exploration for LPW protocols based on data collected using real-world testbeds. 
Specifically, \NAME employs an iterative approach to automatically model a protocol's performance as a function of the exposed parameters based on available experimental data obtained through testbed trials. 
Using an experimentally-derived model, \NAME intelligently selects the next parameter set to be tested using algorithms that minimize the overall number of testbed trials needed to find a solution that meets or is optimal w.r.t. given application requirements\footnote{\highlight{Note that in black-box optimization problems, it is generally impossible to provide a formal proof of global optimality due to the unknown behavior of the objective function. Instead, \NAME provides a structured approach 
balancing exploration and exploitation, which significantly reduces the probability of getting trapped in local minima. By effectively avoiding local minima, the framework increases the likelihood of finding a solution that meets or is optimal for given application requirements with fewer experimental trials.}}.

\fakepar
\NAME's parameter exploration is fully automatic (i.e., users do not need to be actively involved in the process and do not require in-depth understanding of the protocol under study) and the interaction with the testbed infrastructure is seamless (i.e., \NAME can leverage a testbed API to autonomously schedule new tests refining the models capturing the protocol performance as a function of certain parameters). 
To the best of our knowledge, \NAME is the first framework of this kind ever proposed by the LPW community.

\fakepar
After presenting the high-level architecture of the proposed framework~(§\,\ref{sec:overview}), we discuss in detail the design choices of its inner components, navigating through the nuanced trade-offs~(§\,\ref{sec:design_rationale}). 
\highlight{Among others, we compare the use of greedy approaches (often preferred due to their simplicity and low computational demands) to the use of approaches based on Gaussian processes, showing that the latter can effectively handle the inherent noise in experimental data and quickly converge to an optimal parameter set.}
We also enrich \NAME with a way to assess the confidence in the results and estimate their proximity to an optimal solution. 

\fakepar
We implement a concrete instance of the \NAME framework in Python and integrate it with \mbox{D-Cube}~\cite{dcube}, one of the most recent and feature-rich public LPW testbeds. We also open-source our implementation and traces
\footnote{\label{fnote_open}\highlight{The code is available at: \url{http://iti.tugraz.at/APEX}.}} 
to enable the creation of better performing IoT applications and to foster follow-up research on the topic.
We then demonstrate the effectiveness of \NAME experimentally by parametrizing two state-of-the-art multi-hop LPW protocols that have distinct design philosophies for a wide range of application requirements~(§\,\ref{sec:evaluation}). The first protocol, Crystal~\cite{Crystal}, utilizes concurrent transmissions to achieve high network throughput (and has a more deterministic behaviour), while the second protocol, the routing protocol for low-power and lossy networks (RPL)~\cite{rfc6550}, adopts a classical routing approach (and is less deterministic). 
Our results indicate that \NAME can return 
a parameter set that is optimal w.r.t. given application requirements using up to 10.6x less testbed trials 
compared to an exhaustive search of all possible parameter sets, thereby largely minimizing the necessary experimentation time. 
\highlight{We also show the superiority of \NAME over greedy algorithms and approaches based on reinforcement learning, which require, respectively, 4.5x and 3.25x more testbed trials than \NAME to identify the optimal solution.} 

\section{Motivation \& Key Challenges} 
\label{sec:challenges} 

\noindent
Our goal is to build a framework aiding the parametrization of LPW protocols based on the performance observed on a limited number of experimental trials\footnote{\highlight{In this paper, we explicitly focus on \textit{testbed} experiments, as this is often the preferred way to test and debug the performance of LPW protocols using real hardware~\cite{boano_18}. However, as discussed in~§\,\ref{sec:discussion}, the experimental data could also be collected through simulation or in a real-world deployment. \NAME's functionality is, in fact, not bound to the use of LPW testbed facilities.}}. 
Such framework should \highlight{operate automatically (i.e., with minimal user intervention), and should} \textit{confidently} return the best parameter set for a certain protocol that satisfies given application requirements with as few testbed trials as possible (or within a given number of trials).

\fakepar
To exemplify the challenges in designing such a framework, consider an illustrative scenario where we are given a protocol such as Crystal~\cite{Crystal}\footnote{Crystal is a well-known example of LPW protocol based on concurrent transmissions. Crystal works by exploiting Glossy~\cite{Glossy} as a flooding primitive and by splitting floods into transmission and acknowledgment pairs. We refer the reader to the work by Istomin et al.~\cite{Crystal} for further details.}, and are tasked to find the \textit{optimal} set of parameters (i.e., the parameter set that statistically provides the best performance) that allows to minimize the average energy consumption~($E_c$) of all nodes in the network while sustaining an average packet reception ratio (PRR) of at least~65\%. 
For~simplicity, we focus on two parameters of Crystal, as summarized in Tab.\,\ref{tab:protocl}: the transmission~(TX) power used by the radio to send packets ($tx\_power$), and a node's maximum number of transmissions ($n\_tx$) during a Glossy flood. Each of these two parameters can take one out of four possible values\footnote{This results in a total of 16 parameter sets, which may suggest the feasibility of an exhaustive search. However, as we show later in this section, an exhaustive search alone may be insufficient to identify the best parameter set.}.

\fakepar
To design a framework that autonomously solves this task (i.e., that parametrizes Crystal such that energy consumption is minimized while satisfying the given constraint on PRR), we need to \highlight{answer the following questions (Qi)}. 


\begin{table}[!t]
  \centering
  \renewcommand{\arraystretch}{1.35} 
  \begin{tabularx}{\columnwidth/2}{|p{1.35cm}|p{2.55cm}|X|}
   \hline
   \textbf{Protocol}  & \textbf{Parameter}  & \textbf{Possible values} \\
   \hline
   \textit{Crystal} & $tx\_power$ & [-5, -3, -1, 0] dBm \\
   \cline{2-3}
   & $n\_tx$  & [1, 2, 3, 4] \\
   \hline
   \hline
   \multirow{3}{=}{\textit{RPL}} & $max\_link\_metric$ & [16, 32, 64] \\
   \cline{2-3}
   & $DIO\_interval$ & [$2^4, 2^8, 2^{12}, 2^{16} $] ms \\
   \cline{2-3}
   & $Rank\_threshold$  & [4, 8, 12, 16] \\
    \hline
\end{tabularx}
    \vspace{+1.25mm}
    \caption{LPW protocols and parameters considered in this work, and their respective range of possible values.}
    \label{tab:protocl}
    \renewcommand{\arraystretch}{1.00} 
\end{table}

\boldpar{\texttt{(Q1)}: What input does the framework require?} 

\noindent
Firstly, the framework needs to receive input about the protocol under consideration. 
This includes knowledge of the parameters to be optimized, their range of possible values\footnote{Expert users can also (but do not have to) offer recommendations on parameter values that are likely to be effective or ineffective.}, 
how they are exposed within the protocol's code-base, and potential inter-dependencies. 
Secondly, the framework should know how to interface with the testbed where the protocol should be tested, in order to autonomously schedule runs that can shed light on the protocol performance 
for various parameter sets. 
In~this context, any limitation about the testbed's availability (e.g., the maximum number of testbed trials that can be performed) 
should also be provided. 

\fakepar
Next, the framework needs to be informed about the application requirement(s) for which the protocol shall be optimized. 
In the context of LPW systems, application requirements are typically translated into three key metrics related to communication performance: \textit{reliability} (in terms of achieved PRR), \textit{latency}, and 
\textit{energy consumption}~\cite{schuss17competition, foukalas19dependability}\footnote{Note that these requirements are not independent: for example, improving reliability and latency typically entails a higher energy expenditure.}. 

\fakepar
Commonly, such requirements are expressed in the form: 
\begin{equation}
\textit{maximize/minimize } \textit{[metric(s)]} \quad \text{s.t.} \quad \textit{constraint(s) [metric]}.
\label{general_opt_goal}
\end{equation}

\highlight{
i.e., one defines an optimization goal specifying which metric(s) should be maximized or minimized, subject to (s.t.) given constraint(s). 
}
\highlight{In our illustrative scenario, we have considered the following requirement: 
\begin{equation}
\textit{minimize } E_c \quad \text{s.t.} \quad \textit{PRR} \geq 65\%.
\end{equation}
}


\boldpar{\texttt{(Q2)}: How to model protocol performance?} 

\noindent 
Once provided with the necessary input, the framework shall (i)~execute testbed trials to quantitatively assess protocol performance, and (ii)~construct different \textit{models} based on the available experimental observations. 
Such models are instrumental to predict protocol performance for unexplored parameter settings \highlight{and guide the optimization process efficiently. For example, given values for $tx\_power$ and $n\_tx$ as parameters, the models can predict information related to metrics such as energy consumption, PRR, and latency for the Crystal protocol. Instead of randomly selecting parameter sets for evaluation, these models help identifying promising parameter sets.} 
One model is built by fitting the \textit{goal values} derived from experimental observations on a subset of parameter sets, where the goal values are numerical values reflecting the metric(s) that should be maximized or minimized. 
Such model embodies the \textit{goal function}, i.e., it captures the performance of every parameter set in terms of the metric(s) specified in the defined optimization goal. 
Additional models derived by the framework represent the performance of every parameter set for each of the metric(s) given as constraint(s), i.e., for each of the
\textit{constraint metric(s)}. In our illustrative example, the framework would build two models capturing energy consumption and PRR as a function of the employed parameter sets: the former embodies the goal function; the latter models protocol performance for the given constraint metric.  

\fakepar
For instance, the sought framework can initially pick six parameter sets for Crystal\footnote{We assume the user has no hint on good parameter values: the framework then picks 6 random sets (\textit{tx\_power},\textit{n\_tx}): (0,4);\,(0,3);\,(-1,2);\,(-1;3);\,(-5,4);\,(-5,2).}, instruct the testbed to run one trial each, and collect the corresponding performance results in terms of PRR and energy consumption. 
Fig.\,\ref{challenge_all}(a) shows an example of the performance measured in these initial testbed trials: 
Crystal always sustains a PRR higher than 65\%, and consumes between 182.0 and 209.5\,J of energy. 
Based on these results, the framework can approximate the protocol's behaviour in terms of energy consumption (goal value) as a function of \textit{n\_tx} and \textit{tx\_power}, e.g., using polynomial regression, resulting in the model shown in Fig.\,\ref{challenge_all}(b), \highlight{where the green surface represents the goal function}\footnote{\highlight{Note that the framework would also derive the model capturing the PRR as a function of \textit{n\_tx} and \textit{tx\_power}, but we omit this plot for brevity.}}. 

\fakepar
In general, a key trade-off needs to be faced w.r.t. the development of such a model: the framework can adopt simple, pre-defined 
models (e.g., linear \highlight{or polynomial}regression) or more flexible, complex models (e.g., Gaussian processes). The trade-off here involves computational demands and information depth: complex models entail higher computational costs in favor of a greater information depth. Although the sought framework does not need to run on an LPW device, it is crucial to achieve a balance between model depth and computational efficiency, especially with a large number of parameters/values. \highlight{Following the principle of Occam's razor, it is preferable to use simpler models with fewer assumptions, as they are often more flexible and better at capturing complex relationships. This ensures that the framework can adapt to the specific requirements of the optimization process.}

\iftrue 

\begin{figure*}[!t]
    \centering
    \includegraphics[width=\textwidth]{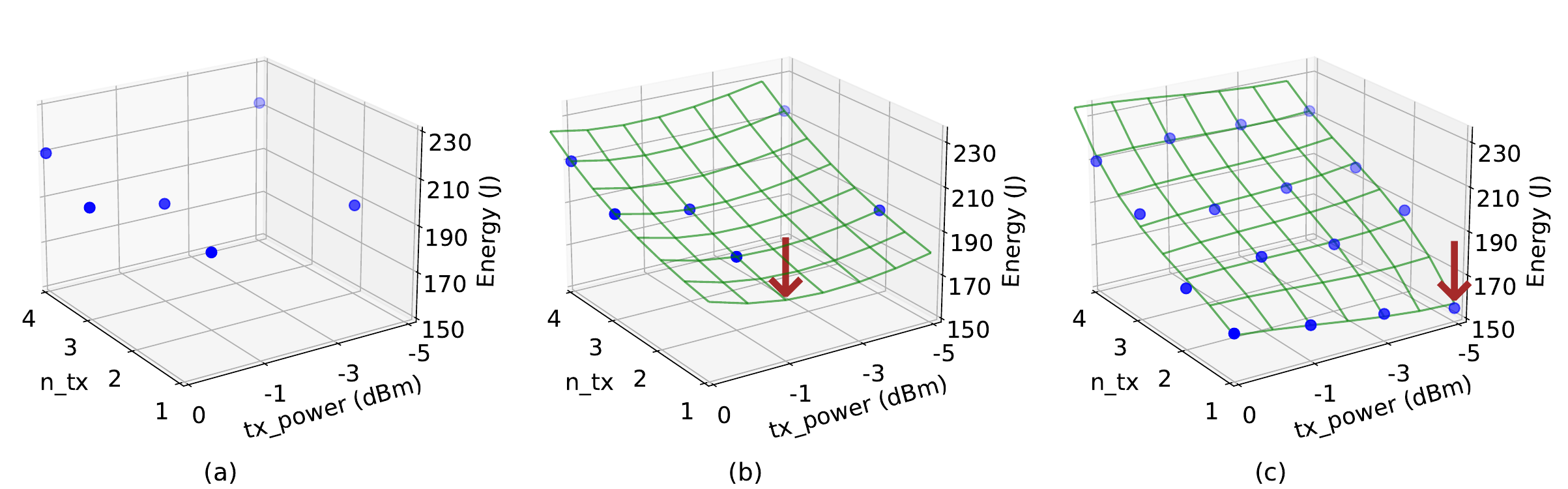}
    \vspace{-7.25mm}
    \caption{Example of parameter exploration and modeling of protocol performance using Crystal. \textmd{Energy consumption experimentally measured through testbed trials using six initial random parameter sets (a); polynomial regression model fitted to these initial experimental results (b); energy consumption and respective fit after an exhaustive search of all sixteen parameter sets (c).}}
    \label{challenge_all}
    \vspace{-0.75mm}
\end{figure*}

\begin{figure*}[!t]
    \centering
    \includegraphics[width=0.475\columnwidth]{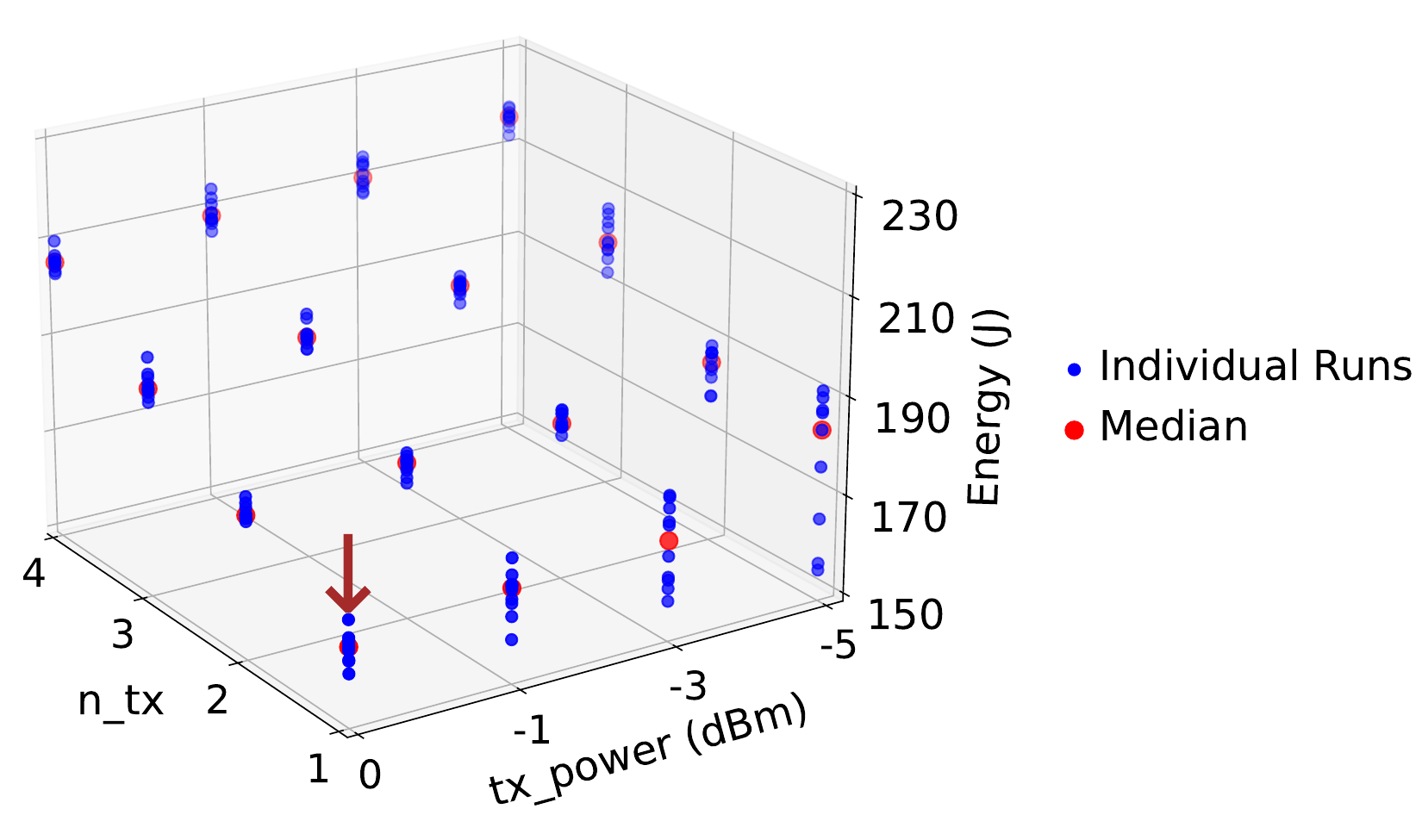}
    \vspace{-3.75mm}
    \caption{Crystal's performance when executing ten experiments per parameter set. \textmd{The best parameter set (brown arrow) is different from that observed in Fig.\,\ref{challenge_all}(c).}}
    \label{Baloo_scatter}
\end{figure*}

\else

\begin{figure*}[!t]
    \centering
    \includegraphics[width=0.84\textwidth]{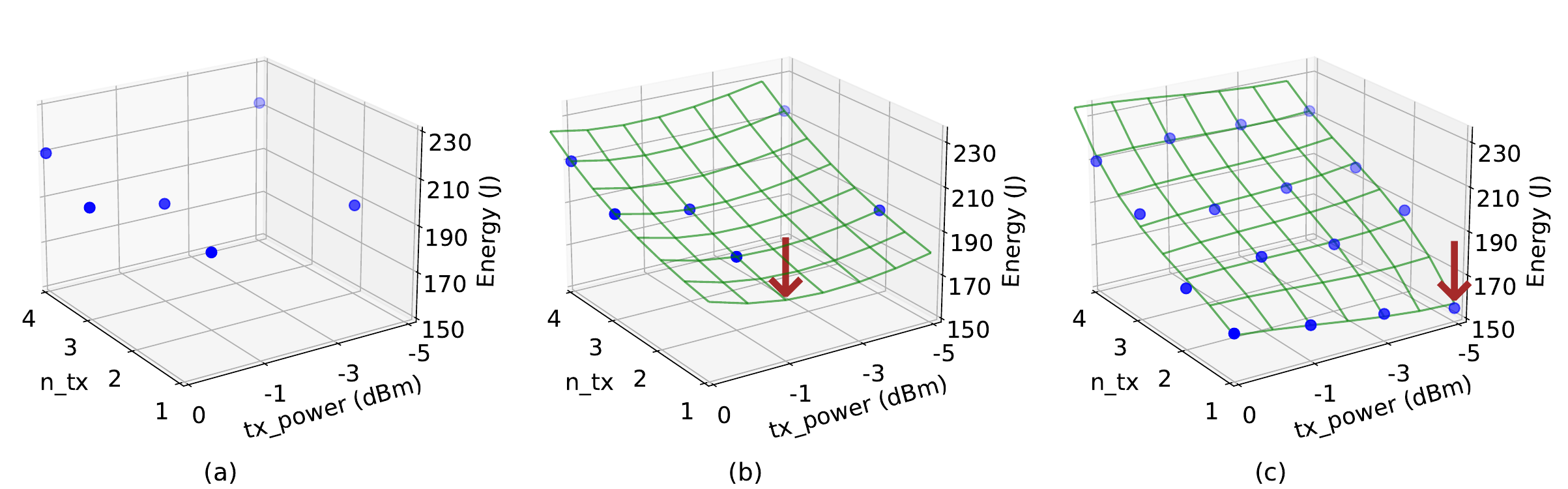}
    \caption{Example of parameter exploration and modelling of protocol performance for Crystal. \textmd{Energy consumption measured in testbed trials using six initial random parameter sets (a); polynomial regression model fitted to these initial results (b); energy consumption and respective fit after an exhaustive search of all sixteen parameter combinations (c).}}
    \label{challenge_all}
\end{figure*}

\fi

\boldpar{\texttt{(Q3)}: How to select the next test-point?}

\noindent
Once fitted models are available, a key challenge is to make an informed selection of the next test-point (i.e., the parameter set whose performance should be measured next): this is essential to guide the optimization process effectively. 
A~straightforward approach might involve the exploitation of the current knowledge to select a test-point that is likely to align best with the application requirements. For example, following the model shown in Fig.\,\ref{challenge_all}(b), we could choose as next test-point the one marked by the brown arrow (i.e., \textit{n\_tx}\,=\,1 and \textit{tx\_power}\,=\,-1\,dBm). 
However, this approach may inadvertently hinder exploration of other potentially better regions within the parameter space\footnote{Overexploitation may also trap the optimization process in local minima.}.
As~can be seen in Fig.\,\ref{challenge_all}(c), which shows the model fitted after an exhaustive search of all sixteen parameter combinations, exploring values with lower TX power would have more quickly led to the optimal parameter set, marked with the brown arrow. 
This example emphasizes the need to balance \textit{exploration} and \textit{exploitation} in surrogate-based optimization problems~\cite{brochu_tutorial_2010}. 

\boldpar{\texttt{(Q4)}: How to handle real-world uncertainties? }

\noindent
Fig.\,\ref{challenge_all}(c) may suggest that the optimal parameter set is \textit{n\_tx}\,=\,1 and \textit{tx\_power}\,=\,-5\,dBm, as marked with the \mbox{brown arrow}. 
In~reality, despite carrying out an exhaustive testing of all parameter combinations\footnote{In this simple example, an exhaustive search requires little effort, as there are only 16 possible sets of parameters. However, exhaustive search would be impractical when dealing with multiple parameters having several possible values each, as the number of experimental trials would rise exponentially.}, this is not necessarily the case: one needs in fact to also account for the \textit{noisiness of experimental data}. 
Fig.\,\ref{Baloo_scatter} shows the results of an exhaustive search when running 10 testbed trials for each parameter combination, with the red dots indicating the median energy for each parameter set. 
The blue dots highlight how the \mbox{energy} measured across individual experiments varies by as much as 36.5\,J for the same parameter set, underlining that multiple experimental iterations are needed to ensure robust parametrization\footnote{When accounting for ten repetitions and using the median value as a reference, the best parameter set satisfying the application requirements is \textit{n\_tx}\,=\,1 and \textit{tx\_power}\,=\,0\,dBm. This may seem counter-intuitive, as the use of a higher \textit{tx\_power} (0\,dBm vs. -5\,dBm) should result in a higher energy consumption. In practice, at startup, Crystal continuously keeps a device's radio active in receiving mode (i.e., without duty cycling), awaiting synchronization with the initiator's first message. The use of a lower \textit{tx\_power} increases the chances to miss synchronization messages, thus resulting in a longer radio active time and a higher energy consumption.}. 
In general, acknowledging and accounting for real-world uncertainties and variations is crucial to increase the statistical robustness of the results (i.e., tolerance to outliers): as outlined by Jacob et al.~\cite{jacob_tool_2021}, ten repetitions are needed in this case to obtain results at the 60th percentile with 99\% confidence. 
This entails a total of 160 experiments, underscoring the impracticality of exhaustive search for a simple case with only 16 parameter sets.

\boldpar{\texttt{(Q5)}: What and when should the framework return?}

\noindent 
The selection of the next test-point based on the fitted models is an iterative process that can in principle run until an exhaustive search is completed. 
As this would imply a significant waste of time and resources (even more so when accounting for repetitions, as highlighted by \textbf{\texttt{Q4}}), the framework should establish a well-defined termination condition. 

\fakepar
Determining when to terminate the optimization process is difficult: stopping the process prematurely may result in sub-optimal outcomes. 
In this regard, the ability to assess the \textit{confidence} in the current solution 
would significantly help in defining the termination condition. 
However, measuring confidence is challenging in noisy scenarios with limited or no prior knowledge of the underlying protocol's behavior. 

\fakepar
Once the decision to stop the optimization process is made, the framework should provide as output the best parameter set discovered during the exploration process. 
The framework should also give hints about the quality of the returned parameter set, e.g., in terms of robustness (confidence that the specified constraints are met) and optimality (confidence that no other parameter combination performs better). 

\vspace{-1.50mm}
\section{\NAME: Overview} 
\label{sec:overview} 
\vspace{-0.50mm}

\begin{figure}[!t]
    \centering
    \includegraphics[width=1.00\columnwidth]{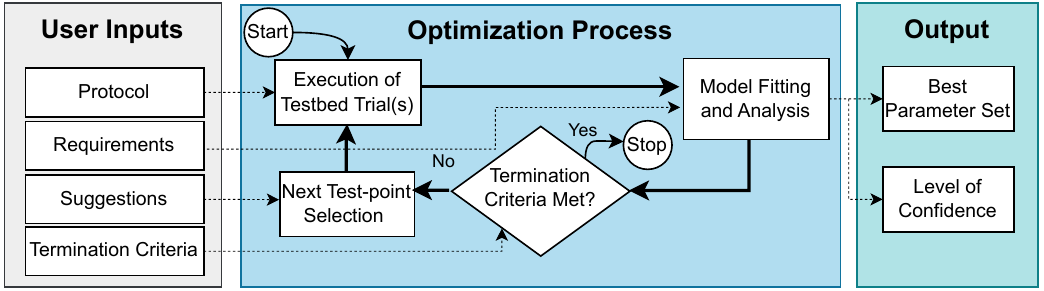}
    \vspace{-7.00mm}
    \caption{High-level overview of the \NAME framework.} 
    \label{framework}
\end{figure}

This section gives an overview of \NAME and its high-level building blocks, which are depicted in Fig.\,\ref{framework}.
As elaborated next, the design of each of these building blocks is guided by the need to address the challenges enumerated in §\,\ref{sec:challenges}. 

\boldpar{User inputs.} 
To tackle the different aspects of \textbf{\texttt{Q1}}, the user needs to provide four types of inputs to \NAME: the protocol under test, application requirements, suggestions for parameter selection (if any), and termination criteria. 
The first input refers to the \mbox{specification} of the protocol being tested (any protocol satisfying the assumptions stated in §\,\ref{subsec: MFA}), and the definition of the parameters to be optimized as well as their possible values. 
The user also needs to supply 
details about the experiments to be performed, such as the duration of individual testbed trials (to meaningfully capture performance), as well as traffic pattern, load, and network topology.

\fakepar
The second input are application requirements expressed as metrics related to communication
performance. 
Each requirement entails one or more constraints \highlight{and an optimization goal (i.e., which metric(s) to maximize or minimize)}, as described in \texttt{\textbf{Q1}}. 
Application requirements are referred to as \textit{$AR$}: those used in this paper are listed in Tab.\,\ref{tab:AR}. 
The sample requirement presented in 
§\,\ref{sec:challenges} (listed as \textit{$AR_1$}) is to minimize the average energy consumption ($E_c$) provided that a minimum PRR of 65\% is sustained, with the considered PRR being the 50$^{th}$ percentile PRR (median).

\fakepar
The third input are suggestions: users with expert knowledge about the protocol can define a favorable parameter space. 
\NAME will account for these suggestions during the optimization. 

\fakepar
The fourth input is the termination condition, which defines the criteria to stop the optimization process. Such criteria can be the maximum number of testbed trials available or the desired level of confidence in the returned parameter set. 

\boldpar{Optimization process.} 
\NAME's parameter exploration starts with an \textit{initial testing phase}, in which the framework \textit{executes testbed trials} for a given number ($n_{init}$) of parameter sets to gain preliminary insights into the protocol's performance, as detailed in~§\,\ref{subsec: ETT}. 
\highlight{If user suggestions are provided in terms of favorable parameter sets, the initial tests prioritize these suggested sets. In the absence of user-suggested parameter sets, or if the number of suggested sets is insufficient, the initial parameter sets can be selected using methods such as Latin hypercube sampling, Sobol sequences\footnote{Latin hypercube sampling is a statistical sampling technique that ensures even coverage of the parameter space, while Sobol sequences provide a low-discrepancy sequence of points for a more uniform sampling distribution.}, or random sampling.} 



\fakepar
After running the initial tests ($n_{init}$ was set to $6$ in the example shown in~§\,\ref{sec:challenges}), the framework proceeds to the \textit{model fitting and analysis} step, which directly addresses \textbf{\texttt{Q2}} and plays a key role in understanding the underlying behavior of the protocol. 
\highlight{In this phase, the framework fits models to the goal values and the values of the respective constraint metric(s), i.e., values of energy consumption and PRR in the case of $AR_1$.} 
Specifically, as detailed in~§\,\ref{subsec: MFA}, \NAME leverages Gaussian processes (GPs) to efficiently model and analyze the limited data available, as they provide a flexible approach for capturing complex relationships without relying on a predefined functional form. 
Moreover, the use of non-parametric models such as GPs also allows \NAME to address real-world uncertainties and thereby address~\textbf{\texttt{Q4}}. 
Then, \NAME identifies the current-best parameter set, i.e., the one with the highest or lowest goal value (depending on the context) that satisfies the specified constraint(s). 
Then, \NAME assesses how confidently this parameter set satisfies the specified constraint(s), e.g., with 98\% confidence, as well as the likelihood that this is the optimal solution. 
These two quantities express \NAME's level of confidence in the current solution.

\fakepar
Afterwards, the framework checks whether the optimization process should continue or whether the given termination criteria have been met. 
\NAME currently tackles \textbf{\texttt{Q5}} by allowing the user to specify as termination criteria the amount of available experimentation time on the testbed (from which the maximum number of testbed trials that can be executed is derived) and/or the level of confidence in the current solution. 
The latter is expressed either as confidence that the specified constraint(s) are met (robustness) or as the likelihood that no better parameter combination exists (optimality). 

\fakepar
If~the termination criteria are not met, the optimization process moves on to the \textit{next test-point selection~(NTS)}, which specifically addresses \textbf{\texttt{Q3}}, by determining the next parameter set to be tested in the next testbed trial. 
The algorithm selecting the next test-point uses the information gained from the fitted model and any user suggestion to select parameter sets that are likely to provide valuable insights into the protocol's behavior or that plausibly lead to better performance. 
§\,\ref{subsec: NTS} will discuss in detail the NTS algorithms proposed for \NAME. 

\boldpar{Output.} 
\NAME's optimization process iterates until the termination criteria are met. 
Upon termination, \NAME returns two outputs (fulfilling \textbf{\texttt{Q5}}). 
First, it provides the best parameter set found during the parameter exploration, i.e., the one that best aligns with the defined metrics and constraint(s). \highlight{As we illustrated in §\,\ref{sec:challenges}, 
considering the statistical measure when determining the best parameter set is crucial. Relying solely on a single value may mislead the optimization process. 
Therefore, in this work, we define the \textit{optimal} (best) parameter set as the one that optimizes the statistical measure over multiple repetitions; in our case, the median, as it is robust to outliers.} 
Second, \NAME generates two confidence metrics: $\alpha$ quantifies the optimality of the returned solution (i.e., the confidence that no other parameter combination performs better than the returned one), whereas $\beta$ expresses its robustness (i.e., the confidence that the given constraint(s) are met). 

\section{\NAME: Key Components}
\label{sec:design_rationale}
Next, we delve into the \NAME framework, exploring its key components. We aim to showcase how \NAME tackles the challenging task of parameter exploration for LPW protocols.

\subsection{Execution of Testbed Trials}
\label{subsec: ETT}
Testbed trials allow to quantify the performance of a protocol. 
A trial typically consists in running a given firmware implementing a network protocol on (a subset of) the testbed nodes, and in retrieving the relevant performance metrics upon the run's completion. 
To this end, one often needs to manually configure the tested firmware (so that it contains the intended parameter set), and to extract the sought performance metrics from raw testbed logs~\cite{Appavoo2019}. 
Note that a few modern tested facilities allow to conveniently compute certain metrics directly in HW (thereby minimizing the user's overhead and enabling a more objective performance evaluation~\cite{adjih15iotlab, lim13flocklab, schuss17competition}), or offer binary patching features to override certain parameter values without the need to regenerate the firmware~\cite{Schub2018}. 

\fakepar 
The parameter set to be tested in a trial refers to a specific combination of values of the protocol parameters to be optimized. 
Let \mbox{$D = \{ \boldsymbol{x}_j \mid j = 1, 2, \ldots, N_p \}$} represent all the parameter sets considered, where $\boldsymbol{x}_j$ is the $j$-th parameter set, $N_p$ is the number of total parameter sets. 
$\boldsymbol{x}_j = [v_1,v_2,..,v_b]$, where $v_q$ is the value of the $q$-th parameter in the respective parameter set, and $b$ is the total number of parameters considered. 
The performance metric returned by the testbed after the n-th testbed trial is denoted as $O_i(n,x_j )$, where $x_j$ is the parameter set tested in the n-th testbed~trial. 
$i \in \{1, 2, ..., N_m\}$ in $O_i$ represents different performance metrics of interest, where $N_m$ denotes the total number of performance metrics considered. 
For simplicity, we use $O(\cdot)$ to refer to the observation related to any performance metrics.


\subsection{Model Fitting and Analysis}
\label{subsec: MFA}
Before delving into the details of model fitting, we outline the key assumptions guiding our approach: 
\begin{itemize}[leftmargin=+.20in]
\item We assume that parameter values can be represented in metric space, allowing for meaningful analysis;
\item We assume that the underlying function representing the performance metric is continuous\highlight{\footnote{\highlight{In most cases, performance metrics align with this assumption. Although some parameters may be defined by discrete values (e.g., transmit power levels of 0 dBm, -1 dBm, -3 dBm, and -5 dBm on specific platforms), their relationship with the performance metric can often be modeled as a continuous function. This is because the underlying behavior of the performance metric generally exhibits a smooth and predictable trend with respect to these parameter changes, allowing for meaningful modeling despite the discrete nature of the parameter values.}}};
\item Without loss of generality, we strictly consider the minimization problem (i.e., we minimize the goal value).
\end{itemize} 

\fakepar
We now extend the mathematical notation to define our system model.
%
Let $f_i: D \rightarrow \mathbb{R}$ represent an unknown continuous function from a compact metric space of parameter sets $D$ to a set of real numbers $\mathbb{R}$. 
Also, $i \in \{1, 2, ..., N_m\}$ in $f_i$ represents different performance metrics that are of interest in the respective optimization process. 
For instance, in $AR_1$, $f_1$ denotes the goal function (i.e., energy consumption) and $f_2$ represents the constraint metric (i.e., PRR). 
For simplicity, we consider a common function $f(\boldsymbol{x})$ to represent any $f_i$. 

\fakepar
After the $n$-th testbed trial, we have $n$ observations returned by the testbed. It is noteworthy that there could be multiple observations for a given parameter set. Then, the task of model fitting is to utilize the current observations and to derive a function $g(\boldsymbol{x})$ that approximates the unknown function $f(\boldsymbol{x})$ to guide the optimization process efficiently. 


\fakepar
The models used to derive $g(\boldsymbol{x})$ can be categorized as parametric or non-parametric. Parametric models, such as linear regression and neural networks, assume specific forms and estimate fixed parameters. 
In contrast, non-parametric models, such as random forests and Gaussian processes~(GPs), do not assume a fixed functional form like parametric models do. 
Instead, they directly learn from data, providing flexibility. 
The choice between parametric and non-parametric models depends on the nature of the problem, data availability, computational constraints, and flexibility. 
GPs offer advantages in our context~\cite{Gramacy2020}, as they excel in data efficiency (thus enabling better predictions with minimal initial data), provide uncertainty estimates for guiding optimization, and facilitate sequential decision-making like Bayesian optimization~\cite{brochu_tutorial_2010}. \highlight{While GPs are computationally demanding, this is not a major concern for \NAME, as the latter is meant to run on a server, and not on the resource-constrained devices. We thus choose GPs to guide \NAME's optimization process efficiently.} 

\boldpar{Gaussian Processes (GPs).}
We model $g(\boldsymbol{x})$ as a stochastic process, so that for any finite set of inputs $\boldsymbol{x}$, the corresponding function values~$g(\boldsymbol{x})$ follow a joint multivariate Gaussian distribution. 
Hence, for any parameter set $\boldsymbol{x}$, the possible corresponding function values $g(\boldsymbol{x})$ are represented as a Gaussian distribution. Thus, mean and variance can characterize the values for each parameter set $\boldsymbol{x}$. 

\fakepar
A GP is fully characterized by its mean function $\mu(\boldsymbol{x})$ and covariance function (kernel function).
For a given parameter set $\boldsymbol{x}$, mean and variance are calculated as in~\cite{brochu_tutorial_2010}: 
\begin{equation}
\mu(\boldsymbol{x}) = \boldsymbol{k}^T \boldsymbol{K}^{-1} \boldsymbol{F^o}
\,\,\,\,\,\,;\,\,\,\,\,\,
\sigma^2(\boldsymbol{x}) = c(\boldsymbol{x}, \boldsymbol{x}) - \boldsymbol{k}^T \boldsymbol{K}^{-1} \boldsymbol{k},
\end{equation}
where $\boldsymbol{k}$ is the vector of covariances between $\boldsymbol{x}$ and the observed parameter sets, $\boldsymbol{K}$ is the covariance (kernel) matrix~of the observed parameter sets, $\boldsymbol{F^o}$ is the vector of observed values of the respective observed parameter sets, and $c(\cdot)$ is the covariance (kernel) function determining the similarity between two inputs. 
Common kernel functions include the radial basis function (RBF) kernel and the Matern kernel, each \mbox{offering various degrees of smoothness and flexibility~\cite{Borovitskiy2021,brochu_tutorial_2010}.}

\fakepar
As the GP model treats $g(\boldsymbol{x})$ as a stochastic process, once the GP is fitted to the given observations, it provides a range of possible functions that could approximate $f(\boldsymbol{x})$. 
This implies that there are multiple potential functions for $g(\boldsymbol{x})$. 
The mean of these functions is denoted by $\mu(\boldsymbol{x})$, which is considered the best estimate among these potential functions. 
Therefore, $\mu(\boldsymbol{x})$ is commonly regarded as the most representative approximation of $f(\boldsymbol{x})$, making it effectively $g(\boldsymbol{x})$.

\fakepar
In addition to $\mu(\boldsymbol{x})$ approximating the behavior of the $f(\boldsymbol{x})$, GPs also provide a tool to calculate the associated uncertainty. 
This aids in guiding the optimization process, specifically in selecting the next test-point (discussed in §\,\ref{subsec: NTS}) and also in approximating the achieved optimality (discussed later in this section). 
The uncertainty associated with a given parameter set $\boldsymbol{x}$ is characterized by the variance $\sigma^2(\boldsymbol{x})$.

\begin{figure}[!t]
  \centering
  \includegraphics[width=\columnwidth]{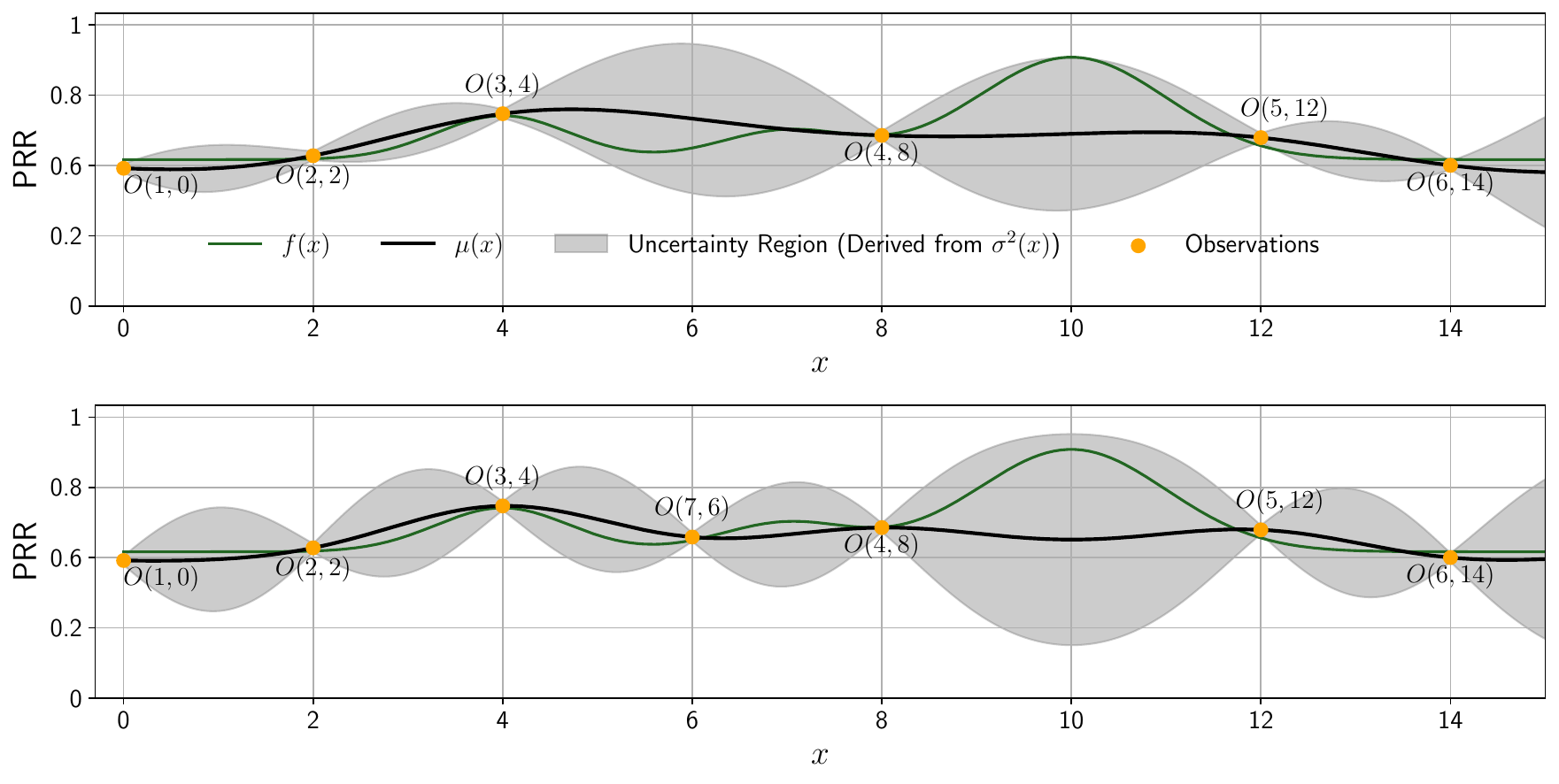}
  \vspace{-7.50mm}
  \caption{Top: GP fit and regression after six testbed trials ($n=6$) for a single parameter ($b=1$), where $x$ represents the values of that parameter. Bottom: updated GP fit and regression after a new trial ($n=7$).}
  \label{fig: SM_2}
\end{figure}

\fakepar
Fig.\,\ref{fig: SM_2}\,(top) shows an example of GP fit after $n=6$ testbed trials. 
Here, $\mu(\boldsymbol{x})$ represent the $g(\boldsymbol{x})$. The uncertainty region around the $\mu(\boldsymbol{x})$, which is depicted as the shaded area around $\mu(\boldsymbol{x})$, is derived from $\sigma^2(\boldsymbol{x})$. 
After executing a new testbed trial ($n=7$), we obtain a new observation at $x=6$, which changes the overall structure of $g(\boldsymbol{x})$ and its related uncertainty, as shown in Fig.\,\ref{fig: SM_2} (bottom). 
The uncertainty near the new observation will decrease, while the uncertainty related to points farther away may remain unchanged or increase. 
This depends on how the uncertainty region is defined, as discussed later in this section under ``\textit{optimality}''. 
In Fig.\,\ref{fig: SM_2}, we can also see that the observation is not always aligned with the underlying unknown function. 
This is because of \textit{noise} in real-world experimental data, which arises from various factors (e.g., changes in environmental conditions, HW-related fluctuations, and RF interference). 
Hence, the observation of the performance in the $n$-th testbed trial can be denoted as:
\begin{equation}
    O(n,\boldsymbol{x}) = f(\boldsymbol{x}) \pm \epsilon (n, \boldsymbol{x}),
\end{equation}
where $\epsilon (n, \boldsymbol{x})$ is the \textit{noise} at the $n$-th testbed trial for the given parameter set $\boldsymbol{x}$.
Once the fitted model is available, \NAME moves to its analysis. 

\boldpar{Best parameter set ($\boldsymbol{x}_n^{+})$.} The analysis begins by identifying the current-best parameter set based on the available results. The filtered parameter sets that satisfy the constraint(s) after the $n$-th testbed trial are $D_n \subseteq D$. 
From these sets, the parameter set with the best goal value (median goal value\footnote{The median is chosen as a simple measure against the outliers.}) is selected as the current-best parameter set. 
The current-best parameter set after the $n$-th testbed trial is denoted as $\boldsymbol{x}_n^{+}$. 
\mbox{Notably}, within the framework's operation, the best parameter set that is returned as output is updated only if the new best set has an equal or greater number of test results compared to the current ones, thus preventing frequent changes and decisions based on outliers.

\boldpar{Robustness ($\beta$).} Then, \NAME will assess how confidently the parameter set satisfies the constraint(s). 
For example, according to $AR_1$ (see Tab.\,\ref{tab:AR}) the median of the PRR should not be smaller than 65\%. 
The confidence metric $\beta$ quantifies how confident \NAME is that the median of PRR is higher than 65\%.
The robustness of a given parameter set $\boldsymbol{x}$ after the $n$-th testbed trial $\beta_n(\boldsymbol{x})$ can be represented through the following inequality using the non-parametric approach~\cite{david_order_2005, jacob_tool_2021}: 
\begin{equation}
\beta_n(\boldsymbol{x}) \geq \mathbb{P}\left(s_l \geq P_p\right)=\sum_{k=0}^{l-1}\left(\begin{array}{l}
N \\
k
\end{array}\right) p^k(1-p)^{N-k},
\end{equation}
where $N$ is the total number of available test results for the parameter set $\boldsymbol{x}$ after the $n$-th testbed trial, $l$ is the number of test results that satisfies the constraint, $s_l$ is the nearest test result to the constraint that satisfies the constraint and $P_p$ is the p-th percentile of the distribution where $p \in (0,1)$ is the percentile. 
In $AR_1$, $p=0.5$ represents the 50-th percentile (median). 
The expression $\mathbb{P}\left(s_l \geq P_p\right)$ provides the confidence that the $p$-th percentile of the respective constraint's metric is above the nearest test result satisfying the constraint for $\boldsymbol{x}$. Thus, we can conservatively take $\beta_n(\boldsymbol{x})$ as $\mathbb{P}\left(s_l \geq P_p\right)$
\footnote{With multiple constraints, the lowest confidence value is chosen as $\beta$.}. 
 
\boldpar{Optimality ($\alpha$).} \NAME computes a second confidence metric $\alpha$ quantifying the level of confidence in the optimality of $\boldsymbol{x}_n^{+}$. 
While this is hard in absence of prior knowledge, 
we can leverage the insights gained from the GPs to derive $\alpha$. 
For that, we initially calculate the GP's lower confidence bound (LCB) for the parameter sets\footnote{In Fig.\,\ref{fig: SM_2}, the LCB is the lower limit of the shaded region.}.  
The LCB at a given parameter set $\boldsymbol{x}$ after the $n$-th testbed trial is given by:
\begin{equation}
\label{Eq_LCB1}
\operatorname{LCB}_n(\boldsymbol{x})=\mu_n(\boldsymbol{x})-\kappa_n \sigma_n(\boldsymbol{x}),
\end{equation}
where $\kappa_n > 0$ is a calibration parameter, $\mu_n(\boldsymbol{x})$ and $\sigma_n(\boldsymbol{x})$ represent the mean and the standard deviation at $\boldsymbol{x}$ after the $n$-th testbed trial. If the bounds are well-calibrated, they accurately reflect the uncertainty in the estimation of the minimum. For a finite parameter space $D$, a well-calibrated $\kappa$ for the $n$-th testbed trial is given as $\kappa_n = \sqrt{2 \log \left(|D| n^2 \pi^2 / 6 \delta\right)}$~\cite{srinivas_2012}, where $\delta \in(0,1)$ represents the strictness of the asymptotic regret bound. The lowest LCB after the $n$-th testbed trial is $\operatorname{LCB}_n^*(\boldsymbol{x}) = \operatorname{min}_{\boldsymbol{x} \in D_n} \operatorname{LCB}_n(\boldsymbol{x})$. Given the bounds are well-calibrated, the instant suboptimality can be approximated as \mbox{$\tau(n) = f(\boldsymbol{x}_n^{+}) - \operatorname{LCB}_n^*(\boldsymbol{x}) $}. 
To capture the relative trend, we calculate the cumulative suboptimality as: 
\begin{equation}
    T(n) = \sum_{t=1}^{n}\tau(t).
\end{equation}

\begin{figure}[t!]
  \centering
  \includegraphics[width=1\columnwidth]{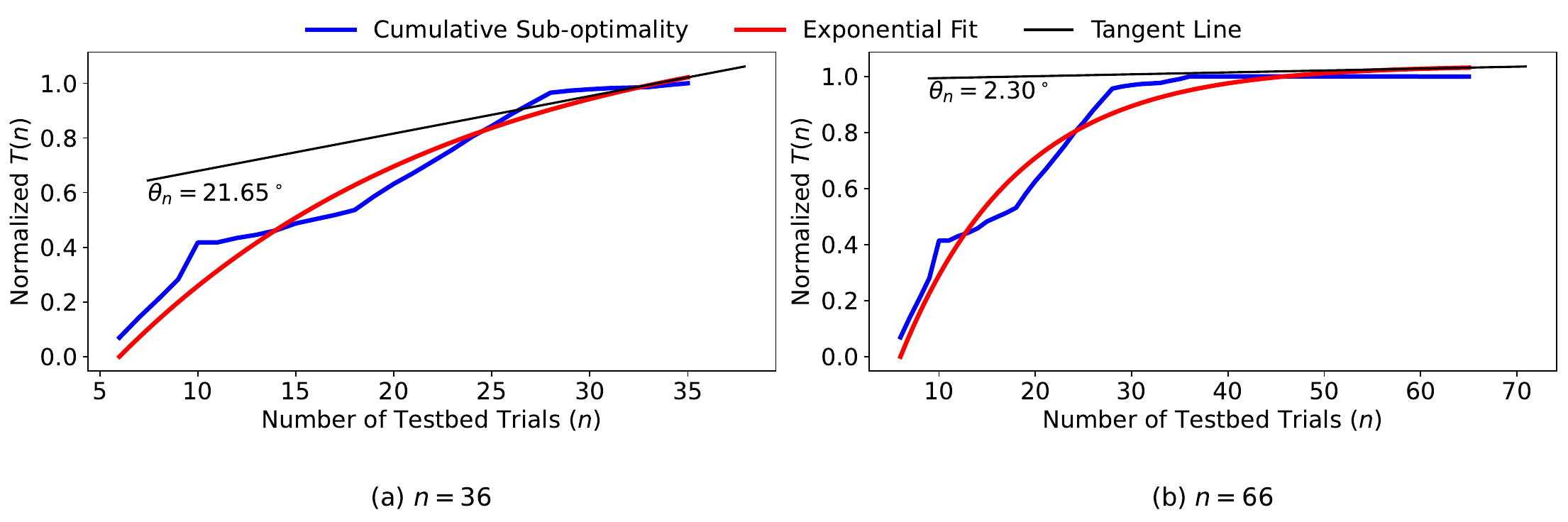}
  \vspace{-7.00mm}
  \caption{\mbox{Illustration of how \NAME derives optimality}. \textmd{We compute an exponential fit of the cumulative sub-optimality $T(n)$, and derive the angle $\theta_n$ of the fit's tangent after the $n$-th testbed trial. A higher $\theta_n$ (e.g., for n=36) indicates a low optimality. After additional testbed trials (e.g., n=66), a low $\theta_n$ indicates closeness to optimality.}}
  \label{fig:CM}
\end{figure}

\noindent
Analyzing $T(\cdot)$ trends aids in understanding optimality progression. Diminishing increments imply nearing optimality; constant or growing increments signal divergence\footnote{\highlight{The phenomenon of diminishing returns implying progress toward optimality generally works even in cases with multiple local minima. However, it may not be effective in scenarios with a sharp loss landscape containing multiple local minima. \NAME's use of simpler functions typically promotes a smoother loss landscape, ensuring that the metric remains effective in most practical cases.}}. Yet, quantifying this trend systematically is complex. We fit an exponential curve to the normalized trend, facilitating derivative calculation and tangent line angle ($\theta_n$) determination at $n$. Fig. \ref{fig:CM} presents an illustrative example of this fit.

\fakepar
A $45^\circ$ angle implies a constant increase in cumulative suboptimality, considered as the worst scenario, thus capped at $45^\circ$ with $\theta_n^\text{'} = \min(45^\circ, \theta_n)$. Then, the optimality confidence metric after the $n$-th testbed trial $\alpha(n)$ is calculated as:
\begin{equation}
  \highlight{\alpha(n)  =  100 (1-\theta_n^\text{'} / 45).}
\end{equation}

\fakepar
Here, $\theta_n^\text{'}$ is in degrees. The $45^\circ$ angle suggests a high likelihood of being far from the optimal, resulting in a confidence level of 0\%. Conversely, angles closer to $0^\circ$ indicate proximity to the optimal, suggesting optimality close to 100\%.

\subsection{Next Test-point Selection (NTS)}
\label{subsec: NTS}

The NTS is crucial because our optimization problem relies on surrogate models~\cite{queipo_surrogate-based_2005}. 
As highlighted by Brochu et al.~\cite{brochu_tutorial_2010}, the core challenge in such problems is balancing exploration and exploitation -- exactly the task of the NTS algorithm. 

\fakepar
The next parameter set that is chosen after the $n$-th testbed trial is $x_{n+1} \in D$. 
Here, we exclusively employ $f(\boldsymbol{x})$ to represent the goal function. Thus, the goal is to find the $\boldsymbol{x}^*$ (optimal parameter set) such that \mbox{$\boldsymbol{x}^*=\operatorname{argmin}_{\boldsymbol{x} \in D_c} f(\boldsymbol{x})$}, where $D_c$ is the actual parameter set that satisfies the constraint(s) to which the optimization goal is subject  (see Eq.~\ref{general_opt_goal}).

\fakepar
We explore two specific NTS algorithms incorporating the insights from GPs, and evaluate their performance in \S\,\ref{sec:evaluation}.

\boldpar{GP-LCB.} The first algorithm focuses on minimizing the lower confidence bound of the estimated model, providing a balanced approach for identifying promising regions likely to contain the optimal solution. The LCB of the parameter sets can be calculated as in Eq.\,(\ref{Eq_LCB1}), where $\kappa_n$ helps to balance the exploration and exploitation.  Then, in the GP-LCB approach, we choose the next test-point as \mbox{$x_{n+1} =\operatorname{argmin}_{\boldsymbol{x} \in D_n} \operatorname{LCB}_n(\boldsymbol{x})$}.

\boldpar{Expected Improvement (EI).} Compared to GP-LCB, which focuses solely on the LCB of the estimated model, this NTS algorithm considers the probability of improvement and the uncertainty over the current-best observation. 
The EI at $\boldsymbol{x}$ after the $n$-th testbed trial is calculated as in~\cite{brochu_tutorial_2010}: 

\begin{equation}
\label{eq:EI}
\mathrm{EI}_n(\boldsymbol{x}) =
\begin{cases}
\left(f(\boldsymbol{x}_n^{+}) - \mu_n(\boldsymbol{x})\right) \Phi(Z_n) + \sigma_n(\boldsymbol{x}) \phi(Z_n), & \text{if } \sigma_n(\boldsymbol{x}) > 0, \\
0, & \text{if } \sigma_n(\boldsymbol{x}) = 0
\end{cases}
\hspace{5em} Z_n = \frac{f(\boldsymbol{x}_n^{+}) - \mu_n(\boldsymbol{x})}{\sigma_n(\boldsymbol{x})}.
\end{equation}

\fakepar where $f(\boldsymbol{x}_n^{+})$ is the current-best observation of the goal value and $Z_n$ is the standard improvement. $\Phi(Z_n)$ and $\phi(Z_n)$ are respectively the cumulative distribution function and probability density function of a standard Gaussian distribution. 
In Eq.\,(\ref{eq:EI}), the term $\left(f(\boldsymbol{x}_n^{+}) - \mu_n(\boldsymbol{x})\right) \Phi(Z_n)$ quantifies the potential improvement over $f(\boldsymbol{x}_n^{+})$ by leveraging the mean prediction of the model and the probability of improvement. Conversely, the term $\sigma_n(\boldsymbol{x}) \phi(Z_n)$ captures the uncertainty in the prediction. Together, these components help balancing exploration and exploitation. 
The next test-point is selected as the one which maximizes the EI, i.e., \mbox{$x_{n+1} =\operatorname{argmax}_{\boldsymbol{x} \in D_n} \mathrm{EI}_n(\boldsymbol{x})$}.

\boldpar{Tackling outliers.} Although GP-LCB and EI effectively handle uncertainty, they are still prone to local minima traps~due to extreme outliers around global/local minima and challenges from noisy measurements in the constraint's metric(s). 
The likelihood of being trapped in a local minimum can be quantified by assessing the uncertainty associated with the selected next parameter set. Low uncertainty implies high confidence, indicating limited potential for new information~\cite{kirschner2018information}. 
Having the uncertainty falling below a predefined lower threshold signals the possibility of being stuck in a local minimum. 
In the EI approach, this likelihood can be quantified by EI itself; in GP-LCB, instead, 
with the coefficient of variation~(CV). 
We hence set the thresholds $EI_{min}$ and $CV_{min}$ based on the $EI$ and $CV$'s maximal values\footnote{Note that \NAME allows expert users to override these values: this can be helpful when measurements are known to have a high/low uncertainty.}. 


\fakepar
To address the issue of extreme outliers around global/local minima, we discard the parameter sets with the highest statistical significance (i.e., with the highest number of test results). We then search for the parameter set that is more likely to offer substantial improvements according to the respective NTS approach. 
To address the challenge posed by noise in the constraint's metric(s), we can use the GP model fitted to such 
metric(s). To generalize, we treat constraint(s) as maximal allowable values. Initially, we calculate the LCB values of the respective constraint's metric, $\operatorname{LCB}^c_n(\boldsymbol{x})$. 

\fakepar
The $\operatorname{LCB}^c_n$ values help to identify parameter sets that are very likely to satisfy the constraint. However, our goal is not only to satisfy the constraint, but to improve the goal value as well. 
The respective improvement in the goal value can be calculated as $I_n(\boldsymbol{x}) = f(\boldsymbol{x}_n^{+})-\mu_n(\boldsymbol{x})$. Then, the combined metric $\Delta_n(x)$ is calculated as:
\begin{equation}
\label{const_imp}
\Delta_n(x) = \frac{\operatorname{LCB}^c_n(\boldsymbol{x})}{f_c^{+}} - \frac{I_n(\boldsymbol{x})}{f(\boldsymbol{x}_n^{+})},
\end{equation}

\fakepar
where $f_c^{+}$ is the current-best observation of the constraint metric \highlight{\footnote{\highlight{$f_c^{+}$ is bounded by the respective constraint's threshold value. For example, if the constraint is specified as "PRR~$\geq$~65\%", the maximum value of $f_c^{+}$ is capped at 65, as only parameter sets that currently fail to meet the constraint are considered.}}.} Both terms are divided by their respective current-best observation to reduce the biases. Then, the next parameter is selected as \mbox{$\boldsymbol{x}_{n+1} = \operatorname{argmin}_{\boldsymbol{x} \in D_n^{'}} \Delta_n(x)$}, where the complementary set $D_n^{'}$ represents the parameter sets that currently do not satisfy the constraint(s)\footnote{If there are multiple constraints, the one with minimum $\Delta_n(x)$ is chosen.}, i.e., \mbox{$D_n^{'} = {i \in D : i \notin D_n}$}. This approach efficiently explores parameter sets that are likely to satisfy the constraint(s) and provide improvements in the goal value. 

\fakepar 
\NAME \, applies the above techniques to tackle outliers in the goal function and \highlight{constraint metric(s)}  \textit{consecutively in each run}, until the optimization process escapes the trap. \highlight{It is important to note that, while we leverage Gaussian processes and well-established acquisition functions, we have tailored these techniques to address the specific challenges of LPW parametrization. In particular, our approach is designed to handle the presence of outliers in the goal function and constraint metric(s) effectively.}

\fakepar
\section{Evaluation}
\label{sec:evaluation}

We implement \highlight{an open-source} prototype of the \NAME framework in Python$^{\ref{fnote_open}}$, with each of the building blocks depicted in Fig.\,\ref{framework} having their own script. 
This modular approach allows users to easily select or replace various components. 
User input (such as protocol and parameters to be optimized, application requirements, and termination criteria) is incorporated into the \NAME's Python scripts through YAML files.

\fakepar
We then evaluate \NAME empirically, analyzing the performance of its optimization process (i.e., 
\highlight{the number of testbed trials required to return the optimal parameter set} -- \S\,\ref{subsec:eval_nts}), the likelihood of returning the optimal parameter set after a fixed number of testbed trials (i.e., the performance after termination -- \S\,\ref{subsubsec: perfterm}), and how well $\alpha$ reflects this likelihood (\S\,\ref{subsubsec: CM}). 

\subsection{Protocols, Metrics, and Methodology}
\label{subsec: metrics_and_meth}

For our evaluation, we use D-Cube~\cite{dcube}, a public testbed that provides a convenient testing environment for LPW systems. 
We focus on a data collection application over a mesh network \highlight{comprising a total of 48 devices supporting IEEE 802.15.4. Among these, five source nodes generate data that is collected by a single destination node.}

\boldpar{Protocols and parameters.} 
We consider two well-known 
LPW protocols: Crystal~\cite{Crystal} and RPL~\cite{rfc6550}. 
Their different nature (Crystal employs concurrent transmissions, whereas RPL adopts a classical routing approach) allows us to assess the versatility of \NAME.
For Crystal, we use the implementation based on Baloo~\cite{Forno2019} running on \texttt{TelosB} devices; for RPL, we employ Contiki-NG's RPL-Lite~\cite{ContikiRPLLite} running on \texttt{nRF52840-DK} devices. 
Each of the five source nodes transmits packets \highlight{periodically} every $1$ and $5$ seconds for Crystal and RPL, respectively. \highlight{This periodicity reflects the one found in common sensor network applications \cite{Lewandowski2021}.}

\fakepar
Tab.\,\ref{tab:protocl} shows the explored parameters for each of the two protocols. 
For Crystal, we consider the two parameters described in \S\,\ref{sec:challenges}\highlight{: $tx\_power$ (transmission power) and $n\_tx$ (maximum number of transmissions)}. 
\highlight{For RPL, we study the $max\_link\_metric$, a parameter used to reject parents with a higher link metric than a given ETX value; the $DIO\_interval$, which is the interval that controls how frequently DODAG Information Object (DIO) messages are sent to disseminate routing and topology information in the network; and the $rank\_threshold$, a parameter used to avoid network instability by only switching to parents having a link quality higher than a given ETX value.} 

\boldpar{Methodology.} 
When evaluating the framework, we need to consider that any approach has the potential to select the optimal parameter set by chance after the first few testbed trials. However, such an outcome does not necessarily imply the inherent superiority of the approach. 
Understanding the true performance of an approach requires a statistically-significant assessment over \textit{numerous iterations}. 
Therefore, in our evaluation, we repeat the optimization process 1000 times. 
However, performing 1000 iterations through testbed experiments is utterly \highlight{impractical, as it would require several years.} 
In fact, a \textit{single iteration} to derive the best parameter set for RPL would require 65 hours of testbed experiments\footnote{For 36 parameter sets, 6 repetitions, and 18-minute runs (see Tab.\,\ref{tab:other_values}).}. 
\highlight{Therefore, we run  a \textit{single iteration} of the optimization process in \mbox{D-Cube} testbed facility, 
in which we test all $N_p$ parameter sets $N_{r}$ times to account for measurement noise, and in which each individual testbed trial takes $\gamma$ minutes. 
The values of these parameters used in our experiments are summarized in Tab.\,\ref{tab:other_values}.}

\fakepar
Then, we use the dataset collected in the testbed to create 1000 iterations of the optimization process. 
\highlight{
In each iteration, when the optimization process needs to test a specific parameter set, it randomly selects one from the $N_{r}$ testbed trials recorded for that parameter set and uses the corresponding result. Each result can only be used once, i.e., if all available results for a specific parameter set are used up, that parameter set is considered exhausted and is no longer a candidate for selection. In such cases, the process will move on to the next best parameter set from the remaining options. However, if the process requests a parameter set where no recorded result is available, or, by default, those values are not assigned to the parameters, it will return the result of the closest available parameter set in the parameter space\footnote{This choice ensures that we can limit the number of experiments to be performed for each parameter set.}}.


\boldpar{Baselines.} 
\highlight{We compare \NAME's performance against five approaches. 
Specifically, we choose three greedy algorithms (for which the same mathematical notations introduced in §\,\ref{sec:design_rationale} are used to describe them) and two approaches based on reinforcement learning (RL) for the NTS. 
These five baselines are summarized as follows (note that the fitted function by the chosen model after the $n$-th testbed trial is $g_n: D \rightarrow \mathbb{R}$):
} 

\begin{itemize}[leftmargin=+.18in]
\item \textit{Greedy for Exploitation (GEL)} focuses on exploitation. After the $n$-th testbed trial, it selects the next parameter set as \mbox{$x_{n+1} =\operatorname{argmin}_{\boldsymbol{x} \in D_n} g_n(\boldsymbol{x})$} or a random one if $D_n = \emptyset$. 
\item \textit{Greedy for exploration (GER)} always explores the parameter space evenly, 
i.e., for each $N_p$ testbed trials, all the parameters sets will be tested once. 
\item \textit{Greedy for Uncertainty (GUC)} balances exploration and exploitation by identifying uncertain regions in the parameter space. After the $n$-th trial, it computes uncertainty metrics $\zeta_n(\boldsymbol{x})$ for $x \in D_n$. Considering only single-hop neighbors, it assigns $-2$ to $\zeta_n(\boldsymbol{x})$ for each test result at $\boldsymbol{x}$ and $-1$ for each test result among the neighbors. It selects a subset $D_u$ with maximum uncertainty, where \mbox{$D_u = {\boldsymbol{x} \in D_n : \zeta_n(\boldsymbol{x}) = \max_{\boldsymbol{y} \in D_n} \zeta_n(\boldsymbol{y})}$}. 
From $D_u$, the next chosen parameter set is $x_{n+1} =\operatorname{argmin}_{\boldsymbol{x} \in D_u} g_n(\boldsymbol{x})$. If $D_u = \emptyset$, a random parameter set is chosen.

\highlight{
\item \textit{Reinforcement Learning (RL-Step)} is inspired by the approach presented in~\cite{Karunanayake2023}. In this method, the agent operates within a Q-learning framework, but with a more constrained approach where it can adjust only one parameter at a time by a single step. The states represent the current parameter combination, and actions involve incrementing, decrementing, or retaining the value of a single parameter. This approach models a more systematic and localized adaptation of the parameter space, with an $\epsilon$-greedy policy used for action selection to balance exploration and exploitation. A penalty is applied for unsatisfied constraints, encouraging the agent to avoid infeasible configurations.}

\highlight{\item \textit{Reinforcement Learning (RL-Any)} builds upon \textit{RL-Step} by relaxing the restriction of adjusting only one parameter at a time. In RL-Any, the agent can adjust multiple parameters simultaneously, making it a more flexible and generalized version than \textit{RL-Step}. This flexibility is designed to align with other approaches, where the agent can move from any parameter set to any other parameter set, enabling broader exploration of the parameter space. As in \textit{RL-Step}, states correspond to the current parameter combination, and actions represent transitions to new parameter configurations. The Q-table maps state-action pairs to expected performance and dynamically expands to accommodate transitions to any parameter combination. This extended flexibility enables the agent to explore the parameter space more efficiently. An $\epsilon$-greedy policy, along with a penalty for unsatisfied constraints, is used in \textit{RL-Any}, similar to \textit{RL-Step}.}

\end{itemize} 

\highlight{
\noindent
Note that, by default, \NAME uses ordinary \textit{least squares regression} for model fitting due to its simplicity, ease of interpretation, and suitability for similar surrogate-based optimization problems~\cite{Balduin2020}. 
In contrast, the two RL approaches employ a Q-table, which is more suitable for these approaches.
}

\fakepar
For the confidence metric approximating optimality, we employ two baseline approaches. The first tracks variance in the best performance after each trial, quantifying confidence:
\begin{equation}
    \alpha_{B_1}(n) = \left (1 - \frac{f(\boldsymbol{x}_n^{+}) - f(\boldsymbol{x}_{n-1}^{+})}{R_g(n))}\right ) \cdot 100,
\end{equation}

\fakepar
where $R_g(n)$ is the current range of the goal values. It calculates the subsequent change in the goal function and normalizes it to the current observed range. This approach assumes that the lesser the variance, the closer the solution is to optimal. However, it is not sensitive to changes in the parameter space ($D$). For instance, a minimal change in the goal value may obscure a significant parameter shift, potentially indicating that the solution is trapped in a local minimum. To address this, we adjust it as follows:
\begin{equation}
    \alpha_{B_2}(n) =\left [\eta  \alpha_{B_1}(n) + \left (1 -\eta  \right )\left (1 - \frac{d(\boldsymbol{x}_n^{+} - \boldsymbol{x}_{n-1}^{+})}{d_{\max}}\right )\right ] \cdot 100,
\end{equation}

\fakepar
where $d(\cdot)$ calculates the normalized Euclidean distance between two points, $d_{max}$ is the maximum Euclidean distance of the parameter space, and $\eta \in(0,1)$ is the balancing factor between both domains.

\begin{figure}[!t]
    \centering
    \includegraphics[width=0.75\columnwidth]{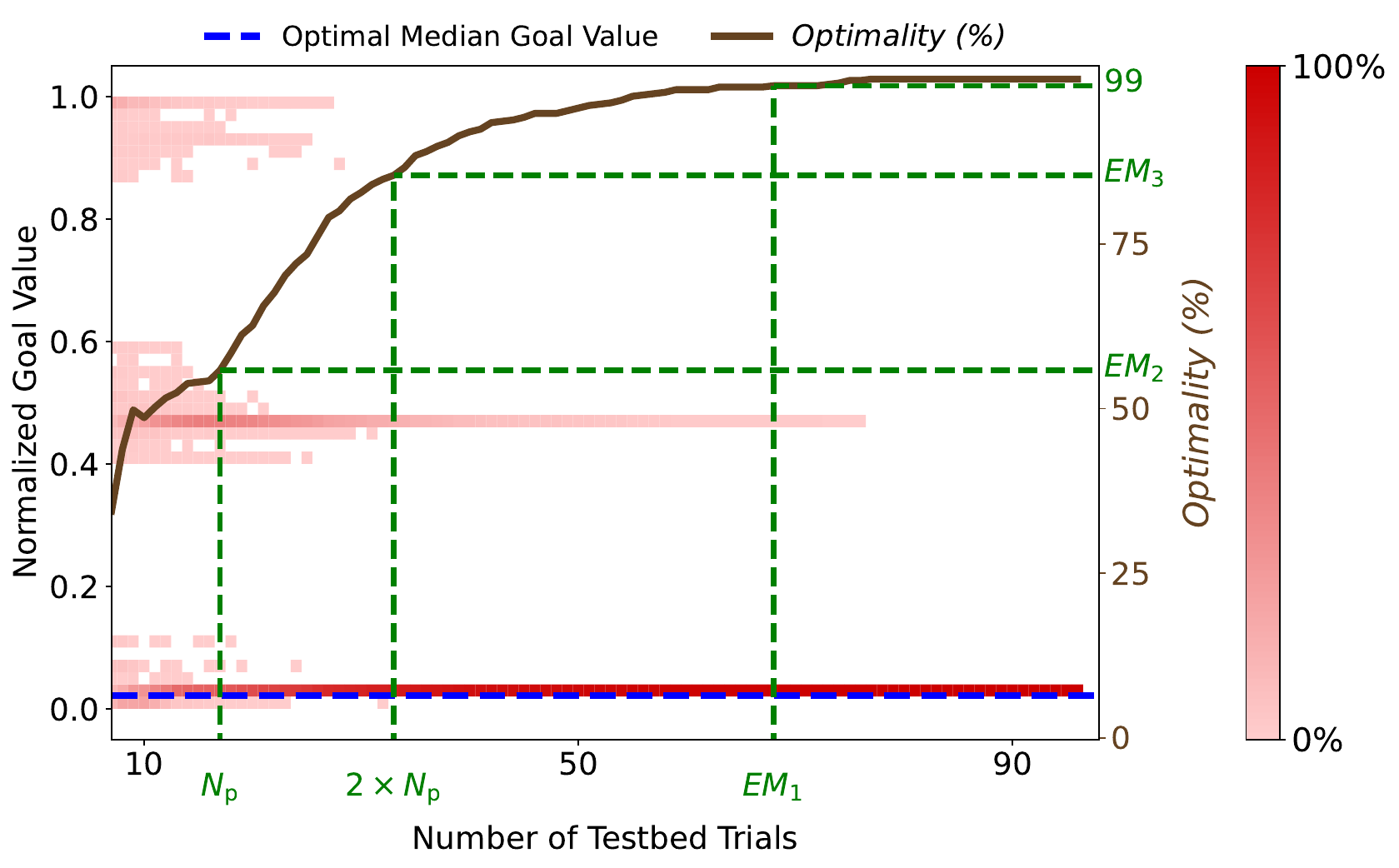} 
    \vspace{-3.75mm}
    \caption{Illustration of a performance evaluation.} 
    \label{fig:heat_map_em}
\end{figure}

\boldpar{Evaluation metrics.} 
Fig.\,\ref{fig:heat_map_em} shows an illustrative example of how we evaluate performance. 
\highlight{This example aims to find the best parameter set for Crystal using GEL for the application requirement ``\mbox{\textit{Minimize $E_c$} | \textit{PRR $\geq$ 96\%''}} (this requirement will later be labeled as $AR_3$, see Tab.\,\ref{tab:AR}).
}
We depict as a heatmap (red color) the distribution of the median goal values returned by 1000 iterations of the optimization process. 
\highlight{Such a heatmap can be used to visualize the points where the optimization process could become trapped, and illustrate how often this occurs before reaching the optimal median goal value (blue dashed line).} 
The solid brown line shows the number \highlight{(in percentage)} of returned parameter sets that is optimal (\textit{optimality}) as a function of the number of executed testbed trials. 
The green dashed lines depict the evaluation metrics. 
The first evaluation metric ($EM_{{1}}$) is defined as the number of testbed trials needed to obtain an \textit{optimality} of 99\%. 
The second evaluation metric ($EM_{{2}}$) is defined as the \textit{optimality} achieved after performing $N_p$ trials. 
Similarly, the third evaluation metric ($EM_{{3}}$) is defined as the \textit{optimality} achieved after performing $2 \cdot N_p$ trials. 

\boldpar{Parameters and application requirements.} 
\highlight{Tab.\,\ref{tab:other_values} lists the relevant parameters used in our evaluation\footnote{Values such as $\delta$ and the RBF length are selected based on recommendations from the literature, and may be overwritten by expert users.}. A single experiment runs for 10 minutes for Crystal and 18 minutes for RPL, with 1 and 3 minutes of settling time, respectively\footnote{\highlight{We have verified with prior experiments that these durations ensure meaningful results.}}. The settling time, which allows the network to stabilize before measurements begin, is particularly necessary for RPL as it is a routing-based protocol and is therefore allocated more time.
Note that users can define the test duration based on prior experience or in alignment with the literature. In the absence of these, the experimental setup can be derived using statistical tools, such as TriScale~\cite{jacob_tool_2021}.
} 

\highlight{
\fakepar 
Tab.\,\ref{tab:AR} lists the application requirements considered in our evaluation. We consider 16 different application requirements (labeled from $AR_1$ to $AR_{16}$) to test how different approaches behave with looser or tighter constraints (e.g., the ability to sustain a minimum PRR as low as 65\% or as high as 97.5\%). 
When selecting the application requirements, we focus on PRR and energy consumption ($E_c$) as performance metrics, and switch their role (as goal value or as constraint metric). Specifically, we pick some application requirements focusing on minimizing $E_c$ while satisfying a given constraint on PRR, and other requirements focusing on maximizing the PRR while satisfying a given constraint on $E_c$. 
}

\highlight{
\fakepar
We specifically use $AR_1$ to $AR_{6}$ to test Crystal, and $AR_7$ to $AR_{12}$ to test RPL, as we aim to maintain a similar range of tightness in the selected constraints, and the two protocols sustain a largely different performance. 
For example, for Crystal, $AR_1$ to $AR_3$ are designed with the goal of reducing $E_c$ with a progressively stricter constraint on PRR. 
$AR_1$ has the loosest constraint, with only one out of sixteen parameter sets failing to meet the required PRR. $AR_2$ introduces a tighter constraint, where half of the parameter sets fail to meet the PRR requirement. $AR_3$ introduces an even stricter constraint, with only three out of sixteen parameter sets satisfying the requirement. $AR_4$ to $AR_6$ follow the same concept, but with the goal of maximizing PRR given a constraint on $E_c$. Also in this case, out of sixteen parameter sets, 15, 8, and 3 satisfy the given constraint, respectively (this can also be observed in Fig.\,\ref{Baloo_scatter}, which shows the performance of different parameter sets as energy drops from 230 to 150\,J). 
We follow the same approach for RPL: $AR_7$ to $AR_9$ focus on reducing $E_c$ while sustaining a minimum PRR, whilst $AR_{10}$ to $AR_{12}$ aim to maximize PRR while not exceeding $E_c$. Also for these application requirements, we strive to maintain a similar level of tightness in the constraints\footnote{\highlight{For RPL, 33, 17 and 7 parameter sets out of 36 satisfy the constraints for $AR_7$ to $AR_9$. Likewise, 32, 18 and 8 parameter sets out of 36 satisfy the constraints for $AR_{10}$ to $AR_{12}$.}}. 

For each of the testbed protocols, we also select two application requirements with very tight constraints, such that only two parameter sets would provide a valid solution: these requirements are labeled $AR_{13}$ to $AR_{14}$ for Crystal, and $AR_{15}$ to $AR_{16}$ for RPL, respectively. These application requirements will be used to demonstrate that \NAME outperforms other approaches also when a very limited number of solutions exists. 
}

\fakepar
Results are expected at the 50th percentile with a maximum confidence of 98\%, which necessitates at least six repetitions ($\rightarrow N_r$\,=\,6) for each parameter set~\cite{jacob_tool_2021}. 
We~set the number of initial trials $n_{init}$ to 6 for 
\NAME and the 
baseline approaches.

\begin{table}
    \centering
    \begin{tabular}{|c|c||c|c||c|c|}
    \hline
    \textbf{Parameter} & \textbf{Value} & \textbf{Parameter} & \textbf{Value} & \textbf{Parameter} & \textbf{Value} \\
    \hline
     $\delta$ & 0.1  & $EI_{min}$  & $EI_{max}/10$ &$\eta$ & 0.5  \\
     \hline
     Kernel for GP & RBF &  $CV_{min}$ & $CV_{max}/10$ & $n_{init}$ & 6 \\
    \hline
    RBF length scale & 1 &  $N_p$ (Crystal)  & 16 &$\gamma$ (Crystal) & 10 minutes \\
    \hline
    $N_r$ & 6 & $N_p$ (RPL)  & 36 & $\gamma$ (RPL) & 18 minutes \\
    \hline
    $\epsilon$ & 0.05 &  &  & &  \\
    \hline
    \end{tabular}
    \renewcommand{\arraystretch}{1.00} 
    \vspace{+0.50mm}
    \caption{Parameter values used in our evaluation.}
    \label{tab:other_values}
\end{table}

\begin{table}[!t]
    \centering
    \begin{tabularx}{\columnwidth}{|c|X|X||c|X|X|}
    \hline
   \textbf{Ref} & \textbf{Goal} & \textbf{Constraint} & \textbf{Ref} & \textbf{Goal} & \textbf{Constraint} \\
    \hline
    $AR_1$ &Minimize $E_c$  & PRR $\geq$ 65\% & $AR_9$ &Minimize $E_c$  & PRR $\geq$ 93\% \\
    \hline
    $AR_2$ & Minimize $E_c$  & PRR $\geq$ 92\% & $AR_{10}$ &Maximize PRR  &$E_c$ $\leq$ 2940 J \\
    \hline     
    $AR_3$ &Minimize $E_c$  & PRR $\geq$ 96\% & $AR_{11}$ &Maximize PRR  & $E_c$ $\leq$ 2885 J \\
    \hline     
    $AR_4$ &Maximize PRR  & $E_c$ $\leq$ 210 J & $AR_{12}$ &Maximize PRR  & $E_c$ $\leq$ 2879 J \\
    \hline 
    $AR_5$ &Maximize PRR  & $E_c$ $\leq$ 190 J & $AR_{13}$ &Minimize $E_c$  & PRR $\geq$ 0.975 \\
    \hline 
    $AR_6$ &Maximize PRR  & $E_c$ $\leq$ 170 J & $AR_{14}$ &Maximize PRR  &$E_c$ $\leq$ 168 J \\
    \hline
    $AR_7$ &Minimize $E_c$  & PRR $\geq$ 65.5\% & $AR_{15}$ &Minimize $E_c$  & PRR $\geq$ 0.947 \\
    \hline
    $AR_8$ & Minimize $E_c$  & PRR $\geq$ 88\% & $AR_{16}$ &Maximize PRR  & $E_c$ $\leq$ 2872 J \\
    \hline 
    \end{tabularx}
    \renewcommand{\arraystretch}{1.00} 
    \caption{Considered application requirements (AR).  
    \textmd{Goal: metric to be improved within the given constraint. 
    Min.: minimize; Max.: maximize; $E_c$: energy consumption.}}
    \label{tab:AR}
\end{table}

\begin{figure*}[!t]
\centering
\includegraphics[width=\textwidth]{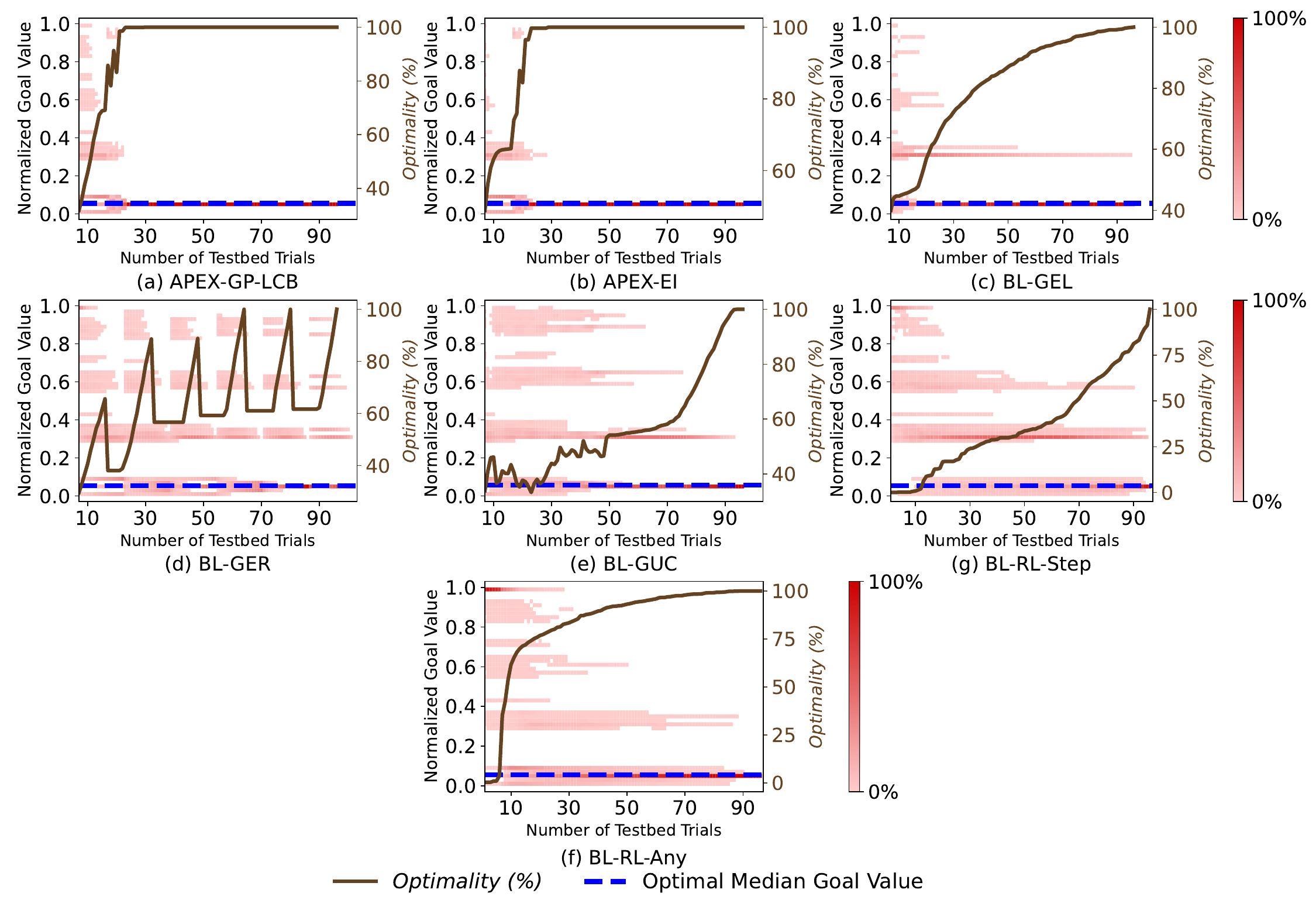}
\vspace{-7.50mm}
\caption{Performance of various NTS algorithms evaluated for $AR_2$. \textmd{\NAME's algorithms are not trapped in local minima and are superior to other baseline approaches.}}
\label{heat_map_all}
\end{figure*}

\subsection{\mbox{Performance of the Optimization Process}}
\label{subsec:eval_nts}


\boldpar{Analyzing the behaviour of NTS approaches.}
We start by evaluating the effectiveness of the next test-point selection (NTS) approaches used by \NAME, showing that they are superior compared to \highlight{the baseline} strategies. 
Fig.\,\ref{heat_map_all} shows the performance of various NTS algorithms when finding Crystal's best parameter set \highlight{for} $AR_2$. 
Whilst \NAME's GP-LCB and EI algorithms are able to quickly escape local minima, this is not the case for the baseline approaches following a greedy strategy (GEL, GER, GUC), \highlight{and for those employing RL. It is to be noted that BL-RL-Any does perform well for a low number of testbed trials; however, its performance becomes similar to that of greedy approaches when it comes to achieving 99\% optimality ($EM_1$).}

\begin{figure}[!t]
    \centering
    \includegraphics[width=\columnwidth]{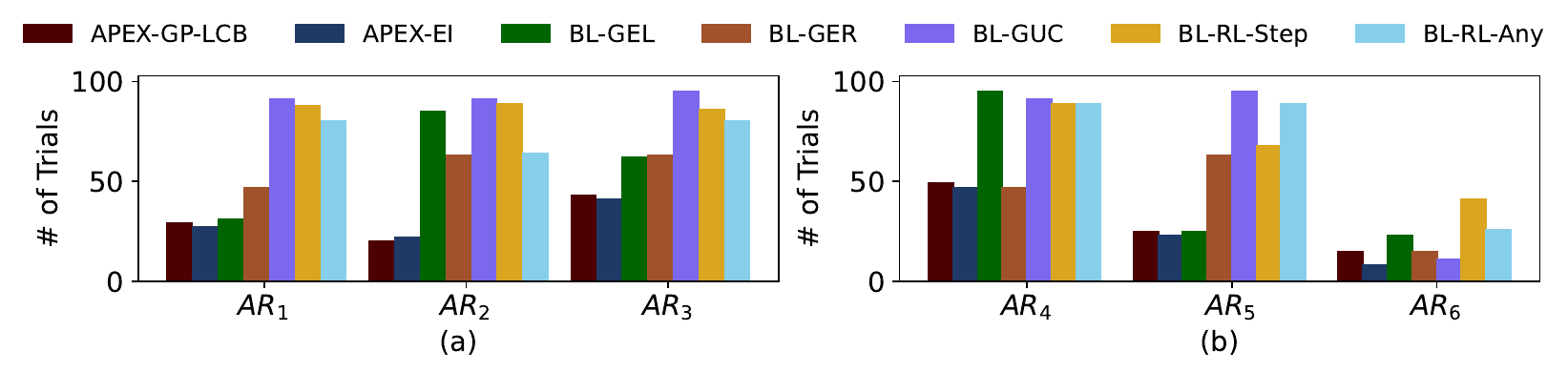} 
    \vspace{-8.25mm}
    \caption{Number of trials needed to obtain an \textit{optimality} of 99\% when evaluating Crystal for different application requirements with varying tightness in their constraint. \textmd{$E_c$ as the goal value \highlight{and the constraint is given as the minimum required PRR }(a), PRR as the goal value \highlight{and the constraint is given as the maximum allowed $E_c$ }(b).}} 
    \label{fig: bar_crystal_EM_1}
\end{figure}

\boldpar{$EM_1$ (Crystal).} 
We analyze next the number of testbed trials needed to achieve an optimality of 99\% when parametrizing Crystal for different application requirements. 
Note that the tightness of constraints varies across different application requirements, with constraints becoming tighter from $AR_1$ to $AR_3$ and from $AR_4$ to $AR_6$.  
Fig.\,\ref{fig: bar_crystal_EM_1} shows that, overall, \NAME consistently outperforms or performs on par with baseline approaches, regardless of the application requirement and the tightness of its constraint.  
For example, for $AR_2$, APEX's GP-LCB and EI need only 20 and 22 trials such that 99\% of the iterations find the optimal parameter set. 
In contrast, BL-GEL, BL-GER, BL-GUC, \highlight{BL-RL-Step, and BL-RL-Any} need 85, 63, 91, \highlight{89, and 64} trials, respectively. 
Hence, \NAME reaches optimality up to \textsc{4.5x} faster than BL approaches.
For Crystal, the traditional exhaustive search approach would need 96 testbed trials to find the optimal parameter set, whereas \NAME requires as little as 9: a \textsc{10.6x} improvement in the best case. 
It is worth noting that the effectiveness of different approaches may vary based on the optimization task's specific requirements. 
If a scenario favors exploitation, BL-GEL would perform better, but in exploration-favoring situations, it may lag. 
Hence, judging an approach's superiority should consider its consistent performance across various situations. 
\textit{\NAME shows consistent performance regardless of the application requirement and the tightness of its constraint.} 

\begin{figure}[!t]
    \centering
    \includegraphics[width=\columnwidth]{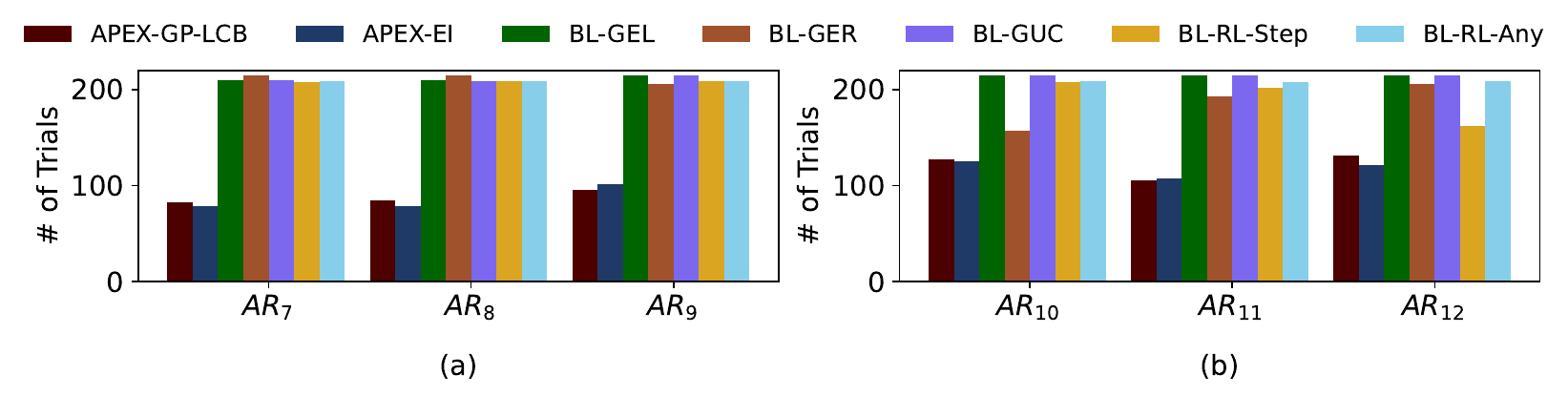} 
    \vspace{-8.25mm}
    \caption{Number of trials needed to obtain an \textit{optimality} of 99\% when evaluating RPL for different application requirements with varying tightness in their constraint. \textmd{$E_c$ as the goal value \highlight{and the constraint is given as the minimum required PRR } (a), PRR as the goal value \highlight{and the constraint is given as the maximum allowed $E_c$ } (b).}} 
    \label{fig: bar_RPL_EM_1}
\end{figure}

\boldpar{$EM_1$ (RPL).} 
We perform a similar evaluation for the RPL protocol, which has a different design philosophy and is less deterministic compared to Crystal. 
Constraints become tighter from $AR_7$ to $AR_9$ and from $AR_{10}$ to $AR_{12}$. 
Fig.\,\ref{fig: bar_RPL_EM_1} shows that \NAME outperforms baseline approaches by a significant margin. This superiority can be attributed to RPL being a less deterministic protocol than Crystal, which emphasizes \NAME's ability to effectively handle noisy measurements. 

\begin{figure}[!t]
    \centering
    \includegraphics[width=1\columnwidth]{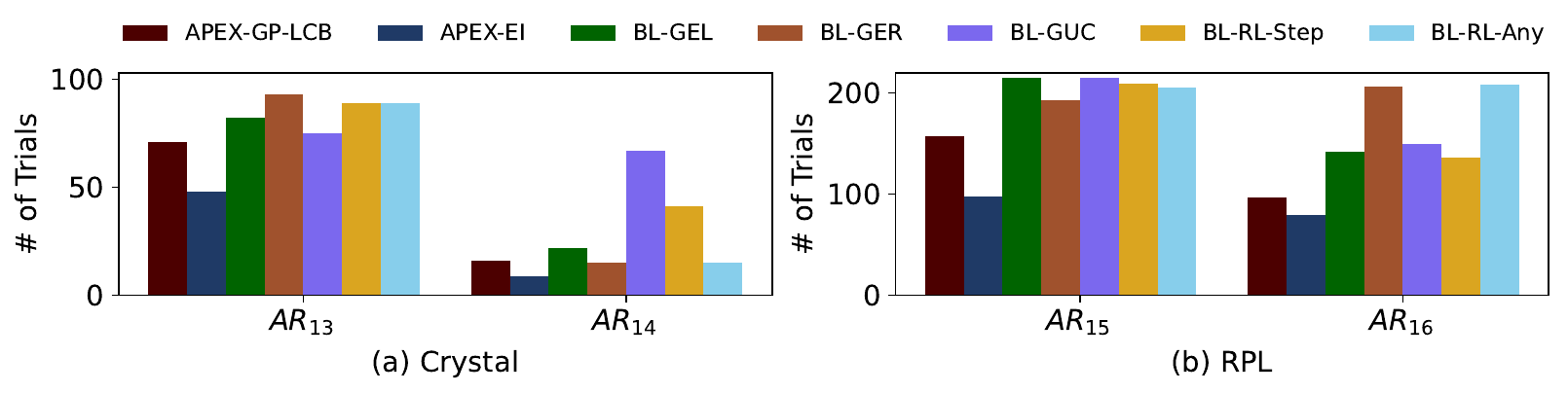} 
    \vspace{-8.25mm}
    \caption{Number of trials required to find~the parameter set that satisfies the constraint ($99$-th percentile).} 
    \label{fig: bar_crystal_both}
\end{figure}

\boldpar{Satisfying \highlight{very tight constraints.}}
\label{subsubsec: Const}
When constraints are exceedingly tight, it may be enough to find at least one parameter set meeting the constraints. 
We hence evaluate how quickly such a parameter set can be found, focusing on \highlight{extreme} cases where only two parameter sets satisfy the constraint.
Fig.\,\ref{fig: bar_crystal_both} shows the number of trials required for 99\% of iterations to discover a parameter set fulfilling given requirements for both Crystal and RPL. 
Observably, \NAME's EI performs better compared to all other approaches, whereas \NAME's GP-LCB performs better or on par with the baseline approaches. \highlight{Specifically, EI requires up to 2.4, 7.4, 2.6, 4.5, and 2.6 times fewer testbed trials to find a parameter set that satisfies the constraints, compared to BL-GEL, BL-GER, BL-GUC, BL-RL-Step, and BL-RL-Any, respectively.}



\noindent
We hence recommend, in general, the use of \NAME's EI rather than GP-LCB, as the former performs better in scenarios with tighter constraint(s). 

\begin{figure}[!t]
    \centering
    \includegraphics[width=\columnwidth]{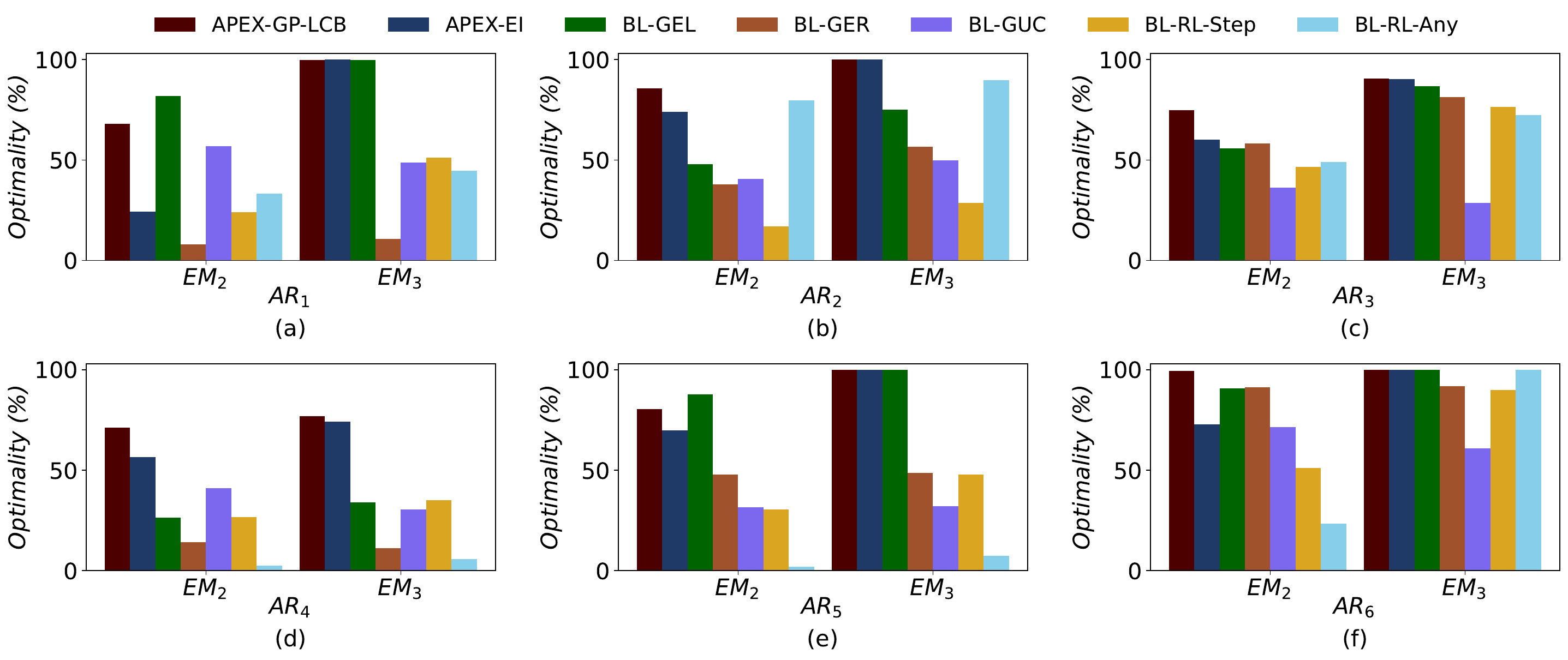} 
    \vspace{-8.25mm}
    \caption{Optimality achieved after a fixed number of trials when evaluating Crystal for application requirements with different tightness of their constraint. \textmd{$E_c$ as the goal value \highlight{with the constraint given as the minimum required PRR} (a)--(c), PRR as the goal value \highlight{with the constraint given as the maximum allowed $E_c$} (d)--(f).}} 
    \label{fig: bar_crystal_EM_2_3}
\end{figure}

\begin{figure}[!t]
    \centering
    \includegraphics[width=\columnwidth]{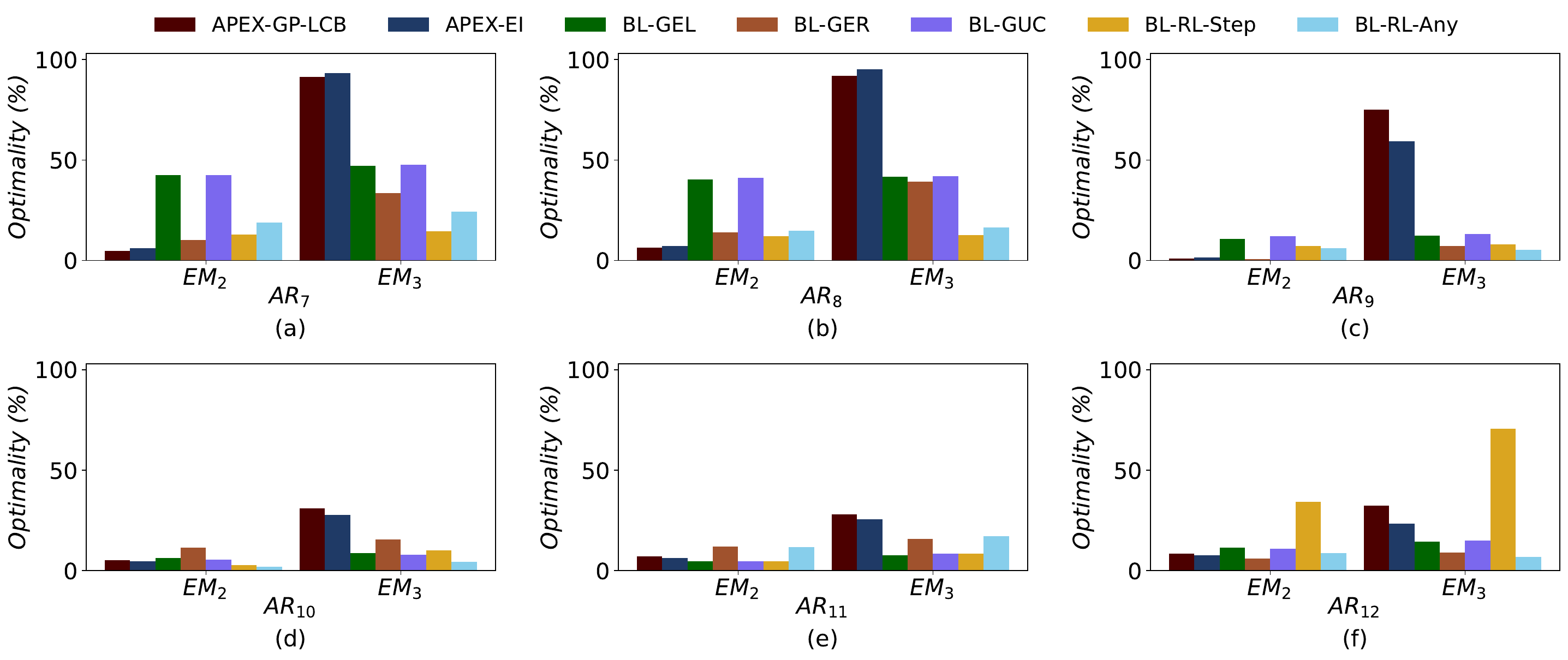} 
    \vspace{-8.25mm}
    \caption{Evaluation of different NTS approaches for the RPL protocol with different tightness of the constraint. The plots show the optimality achieved after a given number of testbed trials using either $E_c$ as the goal value \highlight{with a constraint on the minimum required PRR (a)--(c)} or PRR as the goal value \highlight{with a constraint on the maximum allowed $E_c$} (d)--(f). } 
    \label{fig: bar_RPLEM_2_3}
\end{figure}

\subsection{Performance with a Fixed Number of Trials}
\label{subsubsec: perfterm}

We select the requirements with the loosest to tightest constraints to study the achieved optimality after~a fixed number of trials ($EM_2\,\rightarrow\,N_p$ trials, $EM_3\,\rightarrow\,2\,\cdot\,N_p$ trials).

\boldpar{Crystal.}
Fig.\,\ref{fig: bar_crystal_EM_2_3} shows the results for Crystal (with constraints becoming tighter from $AR_1$ to $AR_3$ and from $AR_4$ to $AR_6$). 
\highlight{
Overall, \NAME outperforms the baseline approaches, especially with a large number of testbed trials ($EM_3$). Still, we can notice that for a few application requirements, some baseline approaches may offer a better/on-par performance when few testbed trials are available. For example, w.r.t. $EM_2$, BL-GEL outperforms \NAME for $AR_1$ and $AR_5$. 
}
This is because \NAME strikes a balance between exploration and exploitation, i.e., it may perform less \mbox{optimally} when very few testbed trials are available ($EM_2$) but catches up over the longer run ($EM_3$).

\boldpar{RPL.} 
Fig.\,\ref{fig: bar_RPLEM_2_3} presents the results for RPL (with constraints becoming tighter from $AR_7$ to $AR_9$ and from $AR_{10}$ to $AR_{12}$). 
\highlight{
As observed for Crystal, \NAME performs very well for $EM_3$ (only for $AR_{12}$, BL-RL-Step performs better\footnote{\highlight{This result highlights a specific case in $AR_{12}$, where limited exploration around the current best, coupled with a penalty for not satisfying the constraint, proves beneficial. In this context, BL-RL-Step outperforms \NAME. Nevertheless, when it comes to achieving 99\% optimality ($EM_1$), BL-RL-Step performs worse compared to the \NAME approaches, as shown in Fig. \ref{fig: bar_RPL_EM_1}.}}). 
However, unlike Crystal, other baselines approaches outperform \NAME for $EM_2$: this is expected, due to RPL's more stochastic nature. 
This overall behavior further reinforces our reasoning for Crystal's results: \NAME may perform sub-optimally with fewer trials due to high noise, requiring more time to balance exploration and exploitation, whereas baseline approaches tend to act greedily, performing better initially, but failing in the long run. 
In fact, in most cases, \NAME significantly outperforms these approaches with respect to $EM_3$, as well as to $EM_1$.}

\highlight{
\boldpar{Summary.} 
Overall, \NAME approaches may perform sub-optimally with fewer trials, especially when there is higher stochasticity. However, they outperform baseline approaches in the long run. It is also noteworthy that \NAME outperforms baseline approaches consistently in finding the optimality in fewer testbed trials ($EM_1$), regardless of the application requirement or the tightness of constraints, which is a key strength for surrogate optimization tasks. In such tasks, where a single approach may excel in one scenario but fail in another, \NAME's ability to maintain strong performance across diverse situations makes it particularly effective.
}

\subsection{Suitability of $\alpha$ to Assess Optimality}
\label{subsubsec: CM}

\begin{figure}[!t]
    \centering
    \includegraphics[width=0.80\columnwidth]{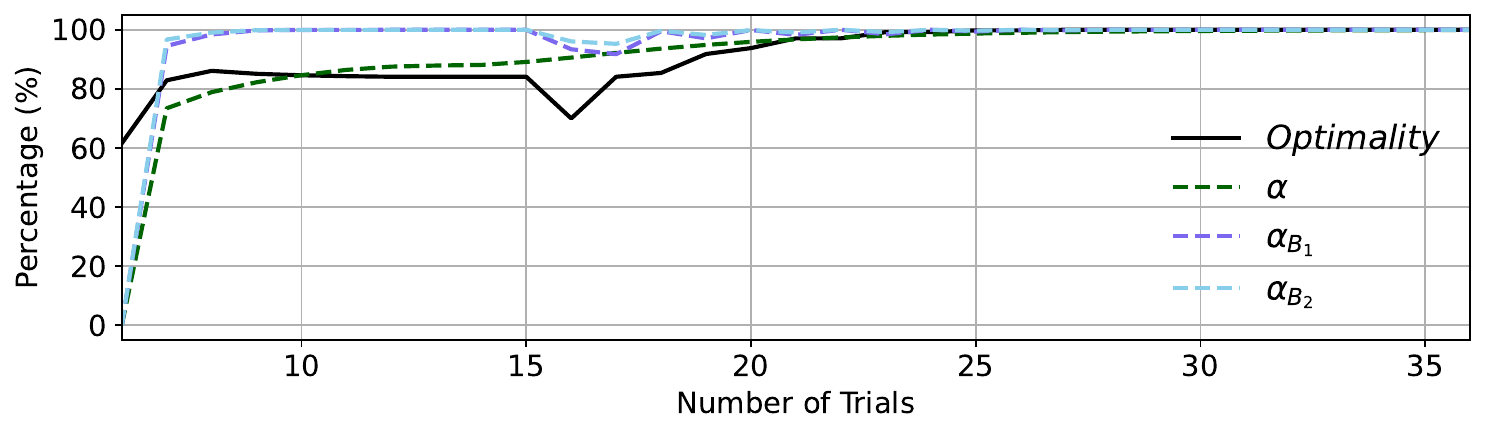} 
    \vspace{-4.25mm}
    \caption{Evaluation of various metrics approximating optimality for a given application requirement ($AR_5$).}
    \label{fig: alpha_comp}
\end{figure}

We now evaluate how well the metric proposed to approximate the optimality achieved ($\alpha$) performs compared to the actual optimality achieved. 
Fig.\,\ref{fig: alpha_comp} illustrates the actual versus predicted optimality for $AR_5$, showing that $\alpha$ tracks the optimality trend better than baseline approaches ($\alpha_{B_1}$ and $\alpha_{B_2}$). 

\fakepar
To provide a quantitative evaluation, we compute the average difference in the number of testbed trials between when the process is terminated using estimated optimality and when the actual required level of optimality is achieved. For example, if a user specifies stopping the process at 80\% confidence in optimality (as estimated by $\alpha$), we calculate how many trials earlier or later the process stops compared to when the actual 80\% optimality is achieved. 
The average of this difference in the number of testbed trials for  $\alpha$, $\alpha_{B_1}$, and $\alpha_{B_2}$ under 80\%, 90\%, and 99\% termination conditions for all application requirements listed in Tab.\,\ref{tab:AR} except $AR_{13}$ - $AR_{16}$\footnote{\label{excluded_ars}We use $AR_1$ to $AR_6$ for Crystal and $AR_7$ to $AR_{12}$ for RPL. $AR_{13}$ to $AR_{16}$ are excluded as they are only intended for identifying a parameter set that satisfies the constraints rather than finding the optimal parameter set.} are 27.9, 52.4, and 52.6 trials, respectively. However, this does not assess how closely the predicted optimality matches the actual achieved optimality. Thus, we compute the root mean square deviation (RMSD) to assess how closely the predicted optimality aligns with the actual achieved optimality. After each testbed trial, we calculate the difference between the predicted and actual optimality, repeating this process for all ARs except $AR_{13}$ to $AR_{16}$\textsuperscript{\ref{excluded_ars}}. The RMSD is then computed based on these differences and averaged across all trials and ARs. The mean RMSD values for $\alpha$, $\alpha_{B_1}$, and $\alpha_{B_2}$ across all requirements are 23.21, 30.9, and 31.2, respectively. These results demonstrate the suitability of \NAME's $\alpha$ as an effective metric for guiding the optimization process relative to the baseline approaches.

\section{Discussion \& Future Work}
\label{sec:discussion}

\boldpar{Multi-objective goal value.} 
In some applications, there~may be multiple metrics that need to be optimized for. For instance, enhancing both PRR and $E_c$ 
could be needed. In such cases, constructing multi-objective goal values requires determining priority weights for metrics and obtaining estimated ranges or expected statistical values for normalization, thus mitigating bias from absolute value differences. If these aspects are available, \NAME can parameterize a multi-objective goal function; otherwise, an alternative approach must be explored to avoid bias favoring one metric over another, pointing to a potential area for future research in understanding Pareto optimality in multi-objective optimization.

\boldpar{Flexible constraint(s).} Here, constraints are defined as single values that must be met with a specified confidence level. However, this approach may discard parameter sets that are close to meeting the constraints but offer better goal values. To address this, understanding how to model flexibility in constraint metrics and integrate it into the parameterization process is an interesting avenue for future investigations. 

\boldpar{Latency.} The time to compute the next test-point, depending on the model and available data, is typically minimal compared to a single trial. For instance, using GPs with 200 input samples, the NTS took about 15 seconds on average\footnote{System specifications: AMD Ryzen 5 PRO 5650U processor with Radeon Graphics, 2.30 GHz; 16.0 GB of RAM (11.8 GB usable). The \texttt{GaussianProcessRegressor} library from sklearn was used for regression.}, a negligible duration compared to 20-minute testbed trials. 
%
However, if latency is an issue, we can pre-calculate future test-points while updating the current model with new \mbox{results} using NTS approaches. 
If the updated model suggests the same point as the one just conducted, we can test the second-best point from the updated model, and so forth.

\boldpar{General applicability \& extensions.}
\highlight{
In this paper, we have explicitly focused on \textit{testbed} experiments. This choice was dictated by the fact that experimentation on testbeds is often the preferred way to test and debug the performance of LPW protocols using real hardware~\cite{boano_18}. 
In fact, testbeds provide a controlled yet realistic environment to explore parameter configurations while approximating field conditions. 
This is very convenient, given the practical challenges and costs of performing extensive parameter optimization directly in a real-world deployment. 
However, \NAME is not bound to testbed experimentation: the experimental data could also be collected through simulation or in a real-world deployment, and the functionality of the framework's inner modules would be agnostic to this (only the way in which the experimental trials should be executed needs to be changed accordingly).  
} 
Moreover, \NAME is designed~for LPW protocol optimization in this study. However, its approach\,of\,treating\,the\,cost\,function\,as\,a black box suggests the potential for extensions to tackle similar challenges in domains where evaluations are expensive and brute force methods with statistical significance are required but impractical. 

\boldpar{Kernel selection.}
For our evaluation, we opted for the RBF kernel (with length scale = 1), a widely-used choice when correlation behavior is unknown. However, users can select different kernels based on preferences for smoothness or other properties. We also assessed the Matérn kernel, another common option, and found its performance comparable.

\highlight{
\boldpar{User in the loop.} 
\NAME is designed to function autonomously once the required user inputs are provided. However, users must make certain decisions to ensure they get the best results from the APEX framework. First, the duration of each experiment and the corresponding experimental settings should be long enough to produce meaningful results. These can be determined based on prior experience, literature, or statistical analysis~\cite{jacob_tool_2021}. Additionally, users should ensure that the selected parameters can be represented in a metric space, meaning they exhibit some form of correlation with changes in their values, even if the relationship is not strictly linear or convex. When dealing with protocols that have only a few parameters, it is possible to optimize all of them. However, when there are many parameters, optimizing all of them may become impractical~\cite{Trunk1979}. In such cases, users can select parameters related to the performance metrics or identify the most influential parameters~\cite{Adan-Lopez2024}, which may require additional experiments. Once the key parameters are chosen, the experimental setup can be determined using tools like TriScale~\cite{jacob_tool_2021}, which suggests appropriate experiment duration and the number of repetitions per parameter set based on specific requirements.}



\section{Related Work} 
\label{sec:related}

\noindent
Over the years, LPW applications have gained significant attention, leading to the development of various LPW protocols tailored to different applications and needs. Optimizing these protocol parameters has become crucial for maximizing performance, prompting extensive research efforts in this domain. Below, we summarize existing works in this field, highlighting important studies, methodologies, and findings.

\boldpar{Parametrization for specific settings.} Early research focused on parameter exploration through measurement studies in specific settings~\cite{Despaux2013, Dong2014, Noda2013}. For example, \cite{Despaux2013} analyzed the impact of duty cycling on average delay and packet loss rate in the IEEE\,802.15.4 protocol stack. \mbox{Similarly},~\cite{Dong2014} proposed a systematic methodology to investigate packet delivery performance in large-scale LPW networks, while~\cite{Noda2013} focused on optimizing packet size and error correction in LPW networks. Efforts also targeted specific protocols like Bluetooth~\cite{Liendo2018, Spork2020}, Zigbee~\cite{Junghee2009}, and LoRa~\cite{KAUR2022107964}. While these studies provided valuable insights, they often focused on individual parameters, limiting inherent trade-offs. 
Joint~multi-parameter exploration began with the work by Fu et al.~\cite{Fu2015}, which provided guidelines for multi-layer parameter optimization across various performance metrics and analyzed a specific protocol in a point-to-point communication system. 
Later, Mazloomi et al.~\cite{Mazloomi2022} have utilized the measurement database from~\cite{Fu2015} to model networks using support vector regression and optimized performance with a genetic algorithm. \highlight{Karunanayake et al.~\cite{Karunanayake2023} have presented an adaptive approach based on reinforcement learning to parametrize the collection tree protocol (CTP~\cite{ctp09gnawali}). Such adaptive approach runs directly on an IEEE~802.15.4 device, i.e., it optimizes protocol parameters on individual nodes without accounting for the actual performance across the entire network. In contrast, \NAME focuses on network-wide optimization. Even when reusing or extending the reinforcement learning method proposed in~\cite{Karunanayake2023}, \NAME yields up to 3.25x better results. Moreover, with \NAME, we provide a generic framework for protocol parametrization and test it on multiple protocols with a large variety of application requirements. 
}

\boldpar{Modular parametrization.} Multiple modular frameworks have been proposed to optimize protocol parameters. One of the pioneering works in this domain is pTUNES~\cite{Zlmmerling2014}, a runtime optimization framework that dynamically adjusts MAC protocol parameters in LPW networks to meet application requirements such as network lifetime, reliability, and latency based on real-time network state data. 
Another framework that expands beyond the optimization of MAC protocol parameters is proposed in~\cite{Oppermann2015} to automate the configuration of IoT communication protocols. However, it requires user expertise in model definition, as it leverages environmental and hardware models for adaptation. The MakeSense project~\cite{Voigt,towards_makesense} has the closest relation to our work, although its primary focus is on reducing costs. 
Specifically, the authors applied reinforcement learning on a simulation of the deployed network, with characteristics of the real network being automatically collected. 
Although impressive, this contribution is not directly applicable to our context, as it would require a new protocol developer to implement MakeSense's low-level functionalities within their protocol. 
In a prior poster abstract~\cite{hydher23exploration}, 
we articulated the necessity for an automated parameter selection framework for LPW systems and outlined its potential design. This paper represents the concrete realization of this initial vision.

\fakepar
Among all these modular frameworks, none enable parametrization for non-experts nor focus on reducing the experimentation time for parametrization. 
With \NAME, we are the first to propose a modular framework that can be used by non-experts while also focusing on reducing the experimentation time.


\section{Conclusion}
\label{sec:conclusion}

In this paper, we introduce \NAME, a framework designed to streamline the parametrization process of LPW protocols. 
By automating parameter exploration based on real-world testbed data and by leveraging Gaussian processes, \NAME offers an efficient solution to protocol parameterization that does not require users to be heavily involved in the optimization process and to possess in-depth protocol understanding. 

\fakepar
Our empirical evaluations underscore the effectiveness of \NAME in efficiently identifying optimal parameter sets, demonstrating a significant reduction in the number of necessary testbed trials compared to traditional approaches. 
Among others, we showcase \NAME's ability to parameterize two state-of-the-art IEEE 802.15.4 protocols, to adapt to diverse application requirements, and to minimize the efforts and costs associated with the parameter tuning process. 

\fakepar 
By making \NAME open-source\footref{fnote_open}, 
we aim to empower researchers and practitioners in the field of LPW systems with the ability to develop dependable networking solutions that can meet stringent application requirements.




\bibliographystyle{ACM-Reference-Format}
\bibliography{references}

\end{document}